\documentclass{article}
\usepackage{ijcai26}

\usepackage{times}
\usepackage{soul}
\usepackage{url}
\usepackage[hidelinks]{hyperref}
\usepackage[utf8]{inputenc}
\usepackage[small]{caption}
\usepackage{graphicx}
\usepackage{amsmath}
\usepackage{amsthm}
\usepackage{booktabs}
\usepackage{algorithm}
\usepackage{algorithmic}
\usepackage[switch]{lineno}
\usepackage[dvipsnames,table,xcdraw]{xcolor}
\usepackage{tcolorbox}
\tcbuselibrary{listings,skins,breakable}
\usepackage{lipsum}               
\usepackage{cuted}
\usepackage{fancyvrb}
\usepackage{multirow}
\usepackage{colortbl}
\usepackage{longtable}
\usepackage{makecell}
\usepackage{tabularx}
\usepackage{enumitem}
\usepackage{adjustbox}
\usepackage{csquotes}
\usepackage{tikz}
\usetikzlibrary{arrows.meta,positioning,fit,backgrounds}
\usepackage{fontawesome5}
\usepackage{listings}

\lstdefinelanguage{json}{
  basicstyle=\ttfamily,
  morestring=[b]",
  morecomment=[l]{//},
  morecomment=[s]{/*}{*/},
}

\title{MindCopilot: Towards Formalizing and Evaluating Granular \\ Human-LLM Co-Writing}

\author{
Youqing Fang$^{1,2}$\textsuperscript{*}
\and
Yinhao Tang$^{1,2}$\textsuperscript{*}
\and
Yanan Sun$^2$\textsuperscript{\dag}
\and
Jiangning Liu$^2$
\and   \\
Ziyi Wang$^2$
\and
Xun Zhao$^2$
\and 
Bin Liu$^1$
\and
Weiming Zhang$^1$
\and   \\
Kuikun Liu$^2$
\and
Wenwei Zhang$^2$
\and
Kai Chen$^2$\textsuperscript{\dag} \\
\affiliations
$^1$University of Science and Technology of China \quad 
$^2$Shanghai AI Laboratory\\
\emails
\{fangyq, tangyinhao\}@mail.ustc.edu.cn,
\{sunyanan, chenkai\}@pjlab.org.cn
}

\begin{document}

\maketitle

\begingroup
\renewcommand{\thefootnote}{}
\footnotetext{* Equal contribution,\, $\dag$ Corresponding author.}
\footnotetext{\url{https://github.com/fangyouqing/MindCopilot}}
\endgroup

\definecolor{prompt}{RGB}{244, 251, 255}
\definecolor{prompt-frame}{RGB}{175, 210, 235}
\definecolor{prompt-light}{RGB}{244, 251, 255}
\definecolor{prompt-light-frame}{RGB}{175, 210, 235}
\definecolor{prompt2}{RGB}{173, 216, 230}
\definecolor{prompt2-frame}{RGB}{0, 51, 102}
\definecolor{prompt3}{RGB}{173, 216, 230}
\definecolor{prompt3-frame}{RGB}{0, 51, 102}
\definecolor{prompt4}{RGB}{173, 216, 230}
\definecolor{prompt4-frame}{RGB}{0, 51, 102}
\definecolor{prompt5}{RGB}{173, 216, 230}
\definecolor{prompt5-frame}{RGB}{0, 51, 102}
\definecolor{contextbgc}{RGB}{242,247,254}    
\definecolor{referencebgc}{RGB}{241,250,245}  
\definecolor{predictionbgc}{RGB}{251,249,244} 
\definecolor{captionbgc}{RGB}{242,242,242}    
\definecolor{contextlabelc}{RGB}{50,100,170}      
\definecolor{referencelabelc}{RGB}{52,122,95}     
\definecolor{predictionlabelc}{RGB}{148,118,78}   
\tcbset{
    HAR/.style={
        enhanced,
        breakable,
        colback=prompt,        
        colframe=prompt-frame,            
        fontupper=\normalsize,               
        boxrule=1pt,                       
        arc=4mm,                           
        left=1mm, right=1mm, top=1mm, bottom=1mm, 
        boxsep=4pt,                        
        before skip=10pt, after skip=10pt, 
        title={\faDatabase~~~\textbf{Prompt: Hierarchical Acceptance Rate}},
        coltitle=black,
        fonttitle=\bfseries\small
    }
}
\tcbset{
    edit_distance/.style={
        enhanced,
        breakable,
        colback=prompt,        
        colframe=prompt-frame,            
        fontupper=\normalsize,               
        boxrule=1pt,                       
        arc=4mm,                           
        left=1mm, right=1mm, top=1mm, bottom=1mm, 
        boxsep=4pt,                        
        before skip=10pt, after skip=10pt, 
        title={\faDatabase~~~\textbf{Prompt: Knowledge-aware Editing Distance}},
        coltitle=black,
        fonttitle=\bfseries\small
    }
}
\tcbset{
    coherence_train/.style={
        enhanced,
        breakable,
        colback=prompt-light,
        colframe=prompt-light-frame,
        fontupper=\normalsize,               
        boxrule=1pt,                       
        arc=4mm,                           
        left=1mm, right=1mm, top=1mm, bottom=1mm, 
        boxsep=4pt,                        
        before skip=10pt, after skip=10pt, 
        title={\faDatabase~~~\textbf{Prompt: Coherence Training Judger}},
        coltitle=black,
        fonttitle=\bfseries\small
    }
}
\tcbset{
    semantic_train/.style={
        enhanced,
        breakable,
        colback=prompt-light,
        colframe=prompt-light-frame,
        fontupper=\normalsize,               
        boxrule=1pt,                       
        arc=4mm,                           
        left=1mm, right=1mm, top=1mm, bottom=1mm, 
        boxsep=4pt,                        
        before skip=10pt, after skip=10pt, 
        title={\faDatabase~~~\textbf{Prompt: Semantics Training Judger}},
        coltitle=black,
        fonttitle=\bfseries\small
    }
}
\tcbset{
    completion_l1/.style={
        enhanced,
        breakable,
        colback=prompt-light,
        colframe=prompt-light-frame,
        fontupper=\normalsize,               
        boxrule=1pt,                       
        arc=4mm,                           
        left=1mm, right=1mm, top=1mm, bottom=1mm, 
        boxsep=4pt,                        
        before skip=10pt, after skip=10pt, 
        title={\faDatabase~~~\textbf{Prompt: Stateless Completion}},
        coltitle=black,
        fonttitle=\bfseries\small
    }
}
\tcbset{
    completion_l2/.style={
        enhanced,
        breakable,
        colback=prompt-light,
        colframe=prompt-light-frame,
        fontupper=\normalsize,               
        boxrule=1pt,                       
        arc=4mm,                           
        left=1mm, right=1mm, top=1mm, bottom=1mm, 
        boxsep=4pt,                        
        before skip=10pt, after skip=10pt, 
        title={\faDatabase~~~\textbf{Prompt: Stateful Completion}},
        coltitle=black,
        fonttitle=\bfseries\small
    }
}

\tcbset{
    logic/.style={
        enhanced,
        breakable,
        colback=prompt,        
        colframe=prompt-frame,            
        fontupper=\normalsize,               
        boxrule=1pt,                       
        arc=4mm,                           
        left=1mm, right=1mm, top=1mm, bottom=1mm, 
        boxsep=4pt,                        
        before skip=10pt, after skip=10pt, 
        title={\faDatabase~~~\textbf{Prompt: Logic}},
        coltitle=black,
        fonttitle=\bfseries\small
    }
}
\tcbset{
    stylistic/.style={
        enhanced,
        breakable,
        colback=prompt,        
        colframe=prompt-frame,            
        fontupper=\normalsize,               
        boxrule=1pt,                       
        arc=4mm,                           
        left=1mm, right=1mm, top=1mm, bottom=1mm, 
        boxsep=4pt,                        
        before skip=10pt, after skip=10pt, 
        title={\faDatabase~~~\textbf{Prompt: Stylistic}},
        coltitle=black,
        fonttitle=\bfseries\small
    }
}
\tcbset{
    semantic/.style={
        enhanced,
        breakable,
        colback=prompt,        
        colframe=prompt-frame,            
        fontupper=\normalsize,               
        boxrule=1pt,                       
        arc=4mm,                           
        left=1mm, right=1mm, top=1mm, bottom=1mm, 
        boxsep=4pt,                        
        before skip=10pt, after skip=10pt, 
        title={\faDatabase~~~\textbf{Prompt: Semantic}},
        coltitle=black,
        fonttitle=\bfseries\small
    }
}

\tcbset{
  holistic/.style={
    enhanced,
    breakable,
    colback=prompt,          
    colframe=prompt-frame,   
    fontupper=\normalsize,
    boxrule=1pt,
    arc=4mm,
    left=1mm, right=1mm, top=1mm, bottom=1mm,
    boxsep=4pt,
    before skip=10pt, after skip=10pt,
    title={\faDatabase~~~\textbf{Prompt: Holistic}},
    coltitle=black,
    fonttitle=\bfseries\small
  }
}

\tcbset{
  captionbg/.style={
    colback=black!5,      
    colframe=black!55,    
    boxrule=0.6pt,
    arc=3pt,
    left=3pt,
    right=3pt,
    top=2pt,
    bottom=2pt,
    boxsep=1pt
  }
}

\tcbset{
  refbox/.style={
    colback=white,
    colframe=black!25,
    boxrule=0.6pt,
    arc=3pt,
    left=3pt,
    right=3pt,
    top=2pt,
    bottom=2pt,
    boxsep=1pt
  }
}

\tcbset{
  refpart/.style={
    enhanced,
    boxrule=0pt,
    frame hidden,
    arc=2pt,
    left=6pt,
    right=6pt,
    top=4pt,
    bottom=4pt
  },
  contextbg/.style={
    refpart,
    colback=contextbgc
  },
  captionbg/.style={
    refpart,
    colback=captionbgc,
    fontupper=\small
  },
  referencebg/.style={
    refpart,
    colback=referencebgc
  },
  predictionbg/.style={
    refpart,
    colback=predictionbgc
  }
}

\tcbset{prompt_info_clarity/.style={
        enhanced,
        breakable,
        colback=prompt,        
        colframe=prompt-frame,            
        fontupper=\normalsize,               
        boxrule=1pt,                       
        arc=4mm,                           
        left=1mm, right=1mm, top=1mm, bottom=1mm, 
        boxsep=4pt,                        
        before skip=10pt, after skip=10pt, 
        title={\faDatabase~~~\textbf{Prompt: Clarity Judge}},
        coltitle=black,
        fonttitle=\bfseries\small
    }
}

\tcbset{prompt_info_content/.style={
        enhanced,
        breakable,
        colback=prompt,        
        colframe=prompt-frame,            
        fontupper=\normalsize,               
        boxrule=1pt,                       
        arc=4mm,                           
        left=1mm, right=1mm, top=1mm, bottom=1mm, 
        boxsep=4pt,                        
        before skip=10pt, after skip=10pt, 
        title={\faDatabase~~~\textbf{Prompt: Content Completeness Judge}},
        coltitle=black,
        fonttitle=\bfseries\small
    }
}

\tcbset{prompt_info_logic/.style={
        enhanced,
        breakable,
        colback=prompt,        
        colframe=prompt-frame,            
        fontupper=\normalsize,               
        boxrule=1pt,                       
        arc=4mm,                           
        left=1mm, right=1mm, top=1mm, bottom=1mm, 
        boxsep=4pt,                        
        before skip=10pt, after skip=10pt, 
        title={\faDatabase~~~\textbf{Prompt: Logical Flow Judge}},
        coltitle=black,
        fonttitle=\bfseries\small
    }
}

\tcbset{prompt_verbatim_question/.style={
        enhanced,
        breakable,
        colback=prompt,        
        colframe=prompt-frame,            
        fontupper=\normalsize,               
        boxrule=1pt,                       
        arc=4mm,                           
        left=1mm, right=1mm, top=1mm, bottom=1mm, 
        boxsep=4pt,                        
        before skip=10pt, after skip=10pt, 
        title={\faDatabase~~~\textbf{Prompt: Generate Verbatim QA}},
        coltitle=black,
        fonttitle=\bfseries\small
    }
}

\tcbset{prompt_interpretive_question/.style={
        enhanced,
        breakable,
        colback=prompt,        
        colframe=prompt-frame,            
        fontupper=\normalsize,               
        boxrule=1pt,                       
        arc=4mm,                           
        left=1mm, right=1mm, top=1mm, bottom=1mm, 
        boxsep=4pt,                        
        before skip=10pt, after skip=10pt, 
        title={\faDatabase~~~\textbf{Prompt: Generate Interpretive QA}},
        coltitle=black,
        fonttitle=\bfseries\small
    }
}

\tcbset{prompt_answer_agent/.style={
        enhanced,
        breakable,
        colback=prompt,        
        colframe=prompt-frame,            
        fontupper=\normalsize,               
        boxrule=1pt,                       
        arc=4mm,                           
        left=1mm, right=1mm, top=1mm, bottom=1mm, 
        boxsep=4pt,                        
        before skip=10pt, after skip=10pt, 
        title={\faDatabase~~~\textbf{Prompt: Answer Questions}},
        coltitle=black,
        fonttitle=\bfseries\small
    }
}

\tcbset{prompt_4o_image/.style={
        enhanced,
        breakable,
        colback=prompt,        
        colframe=prompt-frame,            
        fontupper=\normalsize,               
        boxrule=1pt,                       
        arc=4mm,                           
        left=1mm, right=1mm, top=1mm, bottom=1mm, 
        boxsep=4pt,                        
        before skip=10pt, after skip=10pt, 
        title={\faDatabase~~~\textbf{Prompt: \texttt{4o-Image}}},
        coltitle=black,
        fonttitle=\bfseries\small
    }
}

\tcbset{prompt_owl_4o/.style={
        enhanced,
        breakable,
        colback=prompt,        
        colframe=prompt-frame,            
        fontupper=\normalsize,               
        boxrule=1pt,                       
        arc=4mm,                           
        left=1mm, right=1mm, top=1mm, bottom=1mm, 
        boxsep=4pt,                        
        before skip=10pt, after skip=10pt, 
        title={\faDatabase~~~\textbf{Prompt: \texttt{OWL-4o}}},
        coltitle=black,
        fonttitle=\bfseries\small
    }
}

\tcbset{prompt_4o_html/.style={
        enhanced,
        breakable,
        colback=prompt,        
        colframe=prompt-frame,            
        fontupper=\normalsize,               
        boxrule=1pt,                       
        arc=4mm,                           
        left=1mm, right=1mm, top=1mm, bottom=1mm, 
        boxsep=4pt,                        
        before skip=10pt, after skip=10pt, 
        title={\faDatabase~~~\textbf{Prompt: \texttt{4o-HTML}}},
        coltitle=black,
        fonttitle=\bfseries\small
    }
}

\tcbset{prompt_parser_summarizer/.style={
        enhanced,
        breakable,
        colback=prompt,        
        colframe=prompt-frame,            
        fontupper=\normalsize,               
        boxrule=1pt,                       
        arc=4mm,                           
        left=1mm, right=1mm, top=1mm, bottom=1mm, 
        boxsep=4pt,                        
        before skip=10pt, after skip=10pt, 
        title={\faDatabase~~~\textbf{Prompt: Paper Summarizer}},
        coltitle=black,
        fonttitle=\bfseries\small
    }
}

\tcbset{prompt_parser_filter/.style={
        enhanced,
        breakable,
        colback=prompt,        
        colframe=prompt-frame,            
        fontupper=\normalsize,               
        boxrule=1pt,                       
        arc=4mm,                           
        left=1mm, right=1mm, top=1mm, bottom=1mm, 
        boxsep=4pt,                        
        before skip=10pt, after skip=10pt, 
        title={\faDatabase~~~\textbf{Prompt: Figure Filter}},
        coltitle=black,
        fonttitle=\bfseries\small
    }
}

\tcbset{prompt_planner_matching/.style={
        enhanced,
        breakable,
        colback=prompt,        
        colframe=prompt-frame,            
        fontupper=\normalsize,               
        boxrule=1pt,                       
        arc=4mm,                           
        left=1mm, right=1mm, top=1mm, bottom=1mm, 
        boxsep=4pt,                        
        before skip=10pt, after skip=10pt, 
        title={\faDatabase~~~\textbf{Prompt: Asset Matching}},
        coltitle=black,
        fonttitle=\bfseries\small
    }
}

\tcbset{prompt_planner_painter/.style={
        enhanced,
        breakable,
        colback=prompt,        
        colframe=prompt-frame,            
        fontupper=\normalsize,               
        boxrule=1pt,                       
        arc=4mm,                           
        left=1mm, right=1mm, top=1mm, bottom=1mm, 
        boxsep=4pt,                        
        before skip=10pt, after skip=10pt, 
        title={\faDatabase~~~\textbf{Prompt: Painter}},
        coltitle=black,
        fonttitle=\bfseries\small
    }
}

\tcbset{prompt_planner_commenter/.style={
        enhanced,
        breakable,
        colback=prompt,        
        colframe=prompt-frame,            
        fontupper=\normalsize,               
        boxrule=1pt,                       
        arc=4mm,                           
        left=1mm, right=1mm, top=1mm, bottom=1mm, 
        boxsep=4pt,                        
        before skip=10pt, after skip=10pt, 
        title={\faDatabase~~~\textbf{Prompt: Commenter}},
        coltitle=black,
        fonttitle=\bfseries\small
    }
}

\tcbset{
    vlm_aesthetics_judge/.style={
        enhanced,
        breakable,
        colback=prompt,        
        colframe=prompt-frame,            
        fontupper=\normalsize,               
        boxrule=1pt,                       
        arc=4mm,                           
        left=1mm, right=1mm, top=1mm, bottom=1mm, 
        boxsep=4pt,                        
        before skip=10pt, after skip=10pt, 
        title={\faDatabase~~~\textbf{Prompt: Aesthetics Quality Judge}},
        coltitle=black,
        fonttitle=\bfseries\small
    }
}
\tcbset{
    vlm_element_judge/.style={
        enhanced,
        breakable,
        colback=prompt,        
        colframe=prompt-frame,            
        fontupper=\normalsize,               
        boxrule=1pt,                       
        arc=4mm,                           
        left=1mm, right=1mm, top=1mm, bottom=1mm, 
        boxsep=4pt,                        
        before skip=10pt, after skip=10pt, 
        title={\faDatabase~~~\textbf{Prompt: Element Quality Judge}},
        coltitle=black,
        fonttitle=\bfseries\small
    }
}

\tcbset{
    vlm_layout_judge/.style={
        enhanced,
        breakable,
        colback=prompt,        
        colframe=prompt-frame,            
        fontupper=\normalsize,               
        boxrule=1pt,                       
        arc=4mm,                           
        left=1mm, right=1mm, top=1mm, bottom=1mm, 
        boxsep=4pt,                        
        before skip=10pt, after skip=10pt, 
        title={\faDatabase~~~\textbf{Prompt: Layout Quality Judge}},
        coltitle=black,
        fonttitle=\bfseries\small
    }
}
\begin{abstract}
Recent writing assistants are increasingly shifting from passive, prompt-driven interaction to proactive, suggestion-based completion, which integrates localized continuations into the writing flow and reduces coordination burden. However, existing evaluations simply focus on output quality, failing to capture how users accept, edit, or repair suggestions in real-time interaction, and thus obscuring the true usability of proactive co-writing systems. To address this gap, we adopt a sequential, behavior-centered view of interactive writing and formalize co-writing as a Human-in-the-Loop Markov Decision Process, modeling writing as an interaction shaped by user acceptance and editing decisions. Based on this formulation, we introduce the Co-Writing Fidelity Suite, an interaction-aware metric suite that captures both user–assistant alignment and cognitive editing effort, including Hierarchical Acceptance Rate and Knowledge-aware Editing Distance. We conduct a large-scale simulation study across 16 writing domains, using 1,688 controlled continuation queries sampled from different writing stages. Our analysis reveals systematic effects of interaction structure on acceptance behavior and editing cost. A follow-up user study with 30 participants confirms that these behavioral patterns align with real user experience. Together, our findings demonstrate that interaction-aware evaluation provides insights beyond output-only metrics and informs the design of more effective proactive writing assistants.

\end{abstract}

\section{Introduction}

\label{sec:intro}

Driven by rapid advances in Large Language Models (LLMs)~\cite{liu2025deepseek,comanici2025gemini,singh2025openai}, digital writing workflows have been reshaped, enabling users to draft, revise, and expand text with unprecedented ease~\cite{anthropic_us_usage,openai_how_people_using}.
In most existing systems, writing assistance is realized through \emph{prompt-driven interaction}, where the model remains largely \emph{passive} until explicitly invoked~\cite{bai2022training,pasch2025human}. Authors issue discrete requests, receive complete responses, and manually integrate them into an evolving document. While effective for content generation, this paradigm places the burden of initiative and coordination squarely on the user, offering limited support during the continuous flow of writing~\cite{reza2025co,mysore2025prototypical,dhillon2024shaping,li2026state}.

Recent writing interfaces increasingly adopt a \emph{suggestion-based interactive completion} paradigm~\cite{lee2022coauthor,lee2022coauthor,laban2024beyond}, in which the assistant acts more \emph{proactively} within the writing process. Rather than waiting for prompts, the system surfaces localized continuations directly in the text, and authors iteratively accept, revise, or reject them. This mode of collaboration preserves human agency while reframing writing as a recurrent decision process—one centered on \emph{suggestion review, selective acceptance, and post-hoc editing}. Despite its growing practical relevance, how users behave in such proactive co-writing settings remains poorly understood.

\begin{figure}[t]
\includegraphics[width=0.99\linewidth]{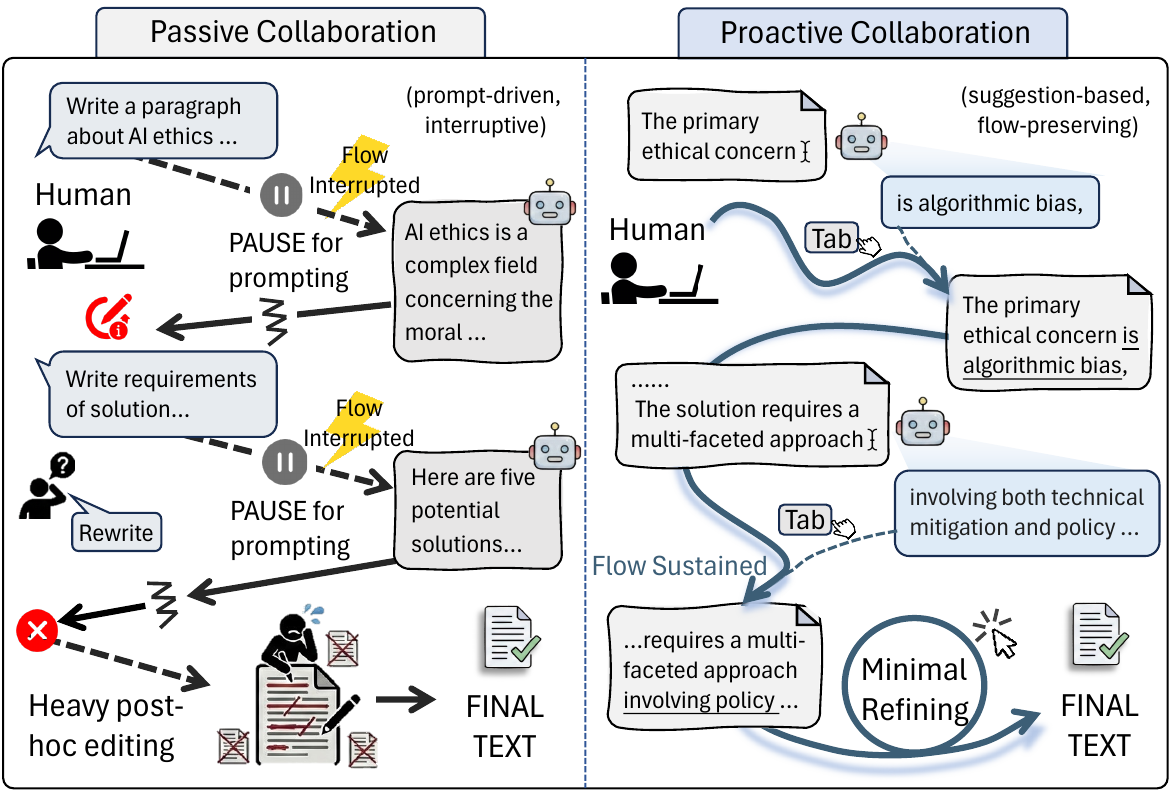} 
  \caption {Two paradigms of LLM-assisted writing.
Prompt-driven interaction interrupts writing flow and shifts coordination burden to the user (left), whereas suggestion-based interaction integrates proactive, localized completions into the writing process, sustaining flow and reducing post-hoc editing (right).}
\end{figure}

This gap is mirrored in prevailing evaluation practices for writing assistants. Widely used paradigms—such as LLM-as-a-Judge or Arena-style comparisons—assess generated text largely in isolation~\cite{chung2025literarytaste,li2025coig,liao2025rlmr,ying2025beyond,fein2025litbench,jia2025writing}, abstracting away the human actions required to integrate suggestions into an ongoing document. As a result, generation quality is often conflated with usability: metrics capture how good a continuation looks, but not whether it aligns with author intent or meaningfully reduces editing effort. Such evaluations fall short for interactive writing, where acceptance behavior and editing cost are as central as surface-level text quality.

To address this limitation, we formalize interactive co-writing as a \emph{Human-in-the-Loop Markov Decision Process (HitL-MDP)}. 
In this formulation, writing is modeled as a sequential decision problem: the assistant generates context-aware suggestions, the user responds through accept/reject/edit actions, and these actions drive state transitions that shape future assistance. 
This abstraction provides a principled framework for analyzing how different interaction structures influence user behavior over time, while admitting a operationalization in which acceptance decisions can be instantiated as a learnable component for downstream optimization.

Building on this framework, we introduce the \textbf{Co-Writing Fidelity Suite}, a set of interaction-aware metrics designed to evaluate human–AI collaboration along two complementary dimensions: alignment and interaction friction. First, we propose \textbf{Hierarchical Acceptance Rate (HAR)}, which treats acceptance not as a binary outcome but as a structured, multi-stage decision, reflecting the cascading filters users implicitly apply during suggestion review. Second, we introduce \textbf{Knowledge-aware Editing Distance (KED)}, which quantifies post-acceptance editing effort by weighting edits according to their cognitive cost, recognizing that correcting knowledge-dense content or factual errors demands substantially more effort than surface-level revisions.

Using these metrics, we conduct a large-scale simulation study across multiple interaction paradigm. We curate a dataset of 60 human-authored articles spanning 16 writing domains and construct 1,688 controlled continuation queries sampled from different writing stages. 
By varying interaction structures while holding content constant, we analyze how factors such as interaction history affect acceptance behavior and editing effort. Guided by insights from these simulations, we further design an adaptive interaction policy and implement a real-world writing interface. A subsequent user study with 30 participants shows that behavioral patterns observed in simulation closely align with subjective user experience.

Our contributions are threefold:
\begin{itemize}
    \item We cast interactive co-writing as a Human-in-the-Loop sequential decision process, providing a principled abstraction for modeling user acceptance, rejection, and editing behaviors over time.
    \item We introduce the Co-Writing Fidelity Suite, a set of interaction-aware metrics—including Hierarchical Acceptance Rate and Cognitive-Weighted Editing Distance—that quantify user–assistant alignment and the cognitive effort required to integrate suggestions.
    \item We conduct a large-scale empirical study combining controlled simulation and real-user experiments, demonstrating that interaction-aware evaluation based on the Co-Writing Fidelity Suite reveals behavioral patterns invisible to output-only metrics.
\end{itemize}

\section{Related Work}

\label{sec:related}

\subsection{Human-AI Co-Writing: From Tool to Collaborative Agency}

The co-writing paradigm is evolving from passive, auxiliary local completion~\cite{lee2022coauthor,coenen2021wordcraft} to mixed-initiative collaboration that frames AI as a \enquote{co-operative partner} preserving user ownership ~\cite{ashkinaze2025ai,zhang2025exploring,he2025contributions,yang2022ai}.
However, evaluation methodologies lag behind. Traditional metrics fail to capture dynamic user experience \cite{shen2023parachute}, while recent attempts at quantifying effort\cite{devatine2024assessing} or keystroke intent \cite{le2025scholawrite} suffer from coarse granularity and rigid definitions. There remains a notable absence of fine-grained, entity-aware metrics to measure cognitive friction in complex interactions.

\subsection{Long-form Writing: Control and Alignment}
With improved context capabilities, generation tasks now encompass document-level long-form content. While works employ outline-driven, recursive strategies to ensure long-range coherence\cite{mirowski2023co,yang2022re3,yang2023doc,bai2408longwriter}—evaluated via coarse-grained dynamic metrics \cite{du2025deepresearch,wu2503writingbench,que2024hellobench}—they predominantly rely on \enquote{black-box} monolithic generation.
This approach leads to \enquote{loss of control} and \enquote{cognitive misalignment}, forcing users into extensive post-generation revisions by ignoring the need for progressive confirmation. The absence of structured interaction protocols prevents the establishment of cognitive consensus during intermediate stages; as a result, even high-quality outputs often diverge from user intent, significantly increasing the burden of post-editing.

\subsection{Creative Evaluation: Quality vs. Intent}
While generative capabilities continue to expand, defining effective creative collaboration in open-domain scenarios remains elusive. Lacking a single gold standard, current approaches depend heavily on preference learning and \enquote{Arena modes} \cite{fein2025litbench,fang2025creation,chung2025literarytaste,li2025coig,jia2025writing}.
Yet, these metrics—rooted in \enquote{win rates} or \enquote{external judging}—reveal a fundamental disconnect: quality does not equal satisfaction. In co-creation, user acceptance is driven less by writing quality than by alignment with the user's specific creative conception. Existing Arena benchmarks fail to capture this alignment with the user's Original Intent. Although AI offers diverse perspectives, it often misses the user's core message, necessitating a paradigm shift from binary \enquote{good/bad} evaluations to the study of \enquote{acceptance and editing behavior patterns}.
\begin{figure*}[t]
  \includegraphics[width=1\linewidth]{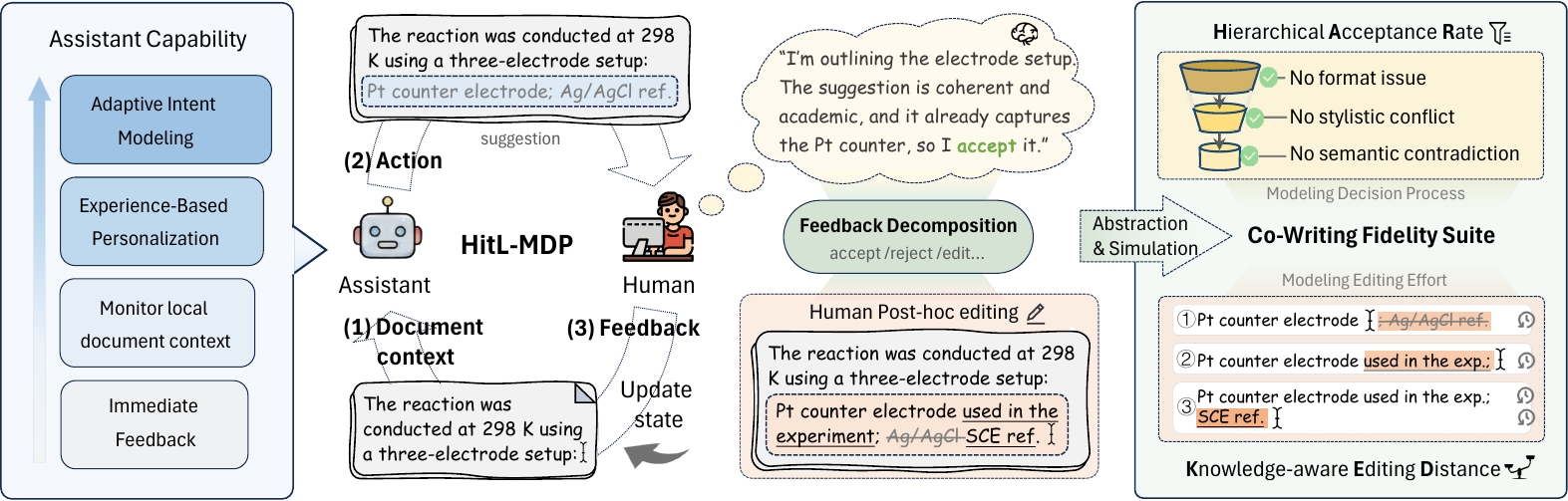} 
  \caption {\textbf{Co-writing as a Human-in-the-Loop decision process.}
The center illustrates co-writing as a HitL-MDP (Sec.\ref{sec:protocol}), where the assistant proposes context-aware suggestions and the human responds actions that update the document state.
Human feedback is decomposed into acceptance followed by post-hoc editing.
The right panel shows the Co-Writing Fidelity Suite (Sec.\ref{sec:fidelity_suite}): HAR models structured acceptance decisions, while KED models editing effort, with darker regions indicating higher cognitive cost for knowledge-dense edits.}

\end{figure*}

\section{Co-Writing Formulation}
\label{sec:protocol}

\subsection{Proactive Collaboration}

We frame interactive auto-completion as a mixed-initiative co-writing setting,
where a human author and a writing assistant collaboratively compose a document
via fine-grained, suggestion-based interactions.
In contrast to prompt-driven generation,
the assistant proactively monitors the evolving document
and offers localized textual continuations during writing,
without requiring explicit user prompts.

Formally, given a document state $s_t$,
the assistant samples a suggestion
\begin{equation}
a_t \sim \pi_\theta(a \mid s_t),
\end{equation}
where $a_t \in \mathcal{A}$ denotes a bounded continuation,
designed to remain local in scope, concise in length, and aligned with the surrounding writing style. 
Crucially, these continuations are non-binding.
The assistant surfaces candidate text,
while all document updates are determined by the human author.

\subsection{Human-in-the-Loop MDP}

Upon receiving a suggestion, the human author provides feedback that governs how the suggestion is treated.
Such feedback span a range of document-editing actions,
including direct adoption, user-mediated modification, explicit dismissal,
and other interactions that alter the proposal’s realization.
Collectively, these forms of feedback are drawn from a structured yet open-ended
user response space $\mathcal{U}$,
and determine whether and how the suggested content contributes
to the subsequent document state.

This asymmetric interaction induces a recurrent decision-making loop.
At step $t$, the assistant observes the current document state $s_t \in \mathcal{S}$
and generates a suggestion $a_t \in \mathcal{A}$.
The human author then responds with a form of feedback $u_t \in \mathcal{U}$,
which specifies how the suggestion is handled.
The document is updated accordingly, yielding a new state $s_{t+1}$,
upon which the assistant conditions to produce the next suggestion.
Over time, document evolution emerges from the repeated interplay between assistant suggestions and human feedback.

We model this iterative process as a Human-in-the-Loop Markov Decision Process (HitL-MDP),
which makes explicit the division of roles between proposing and deciding.
Formally, the process is defined as
$\mathcal{M}=\langle\mathcal{S},\mathcal{A},\mathcal{U},\mathcal{T}\rangle$.
At each step, given the current state $s_t$,
the assistant samples a suggestion according to
$a_t \sim \pi_\theta(\cdot \mid s_t)$,
after which the human selects a form of feedback $u_t$.
State transitions are governed by the transition function
\begin{equation}
s_{t+1} = \mathcal{T}(s_t, a_t, u_t),
\end{equation}
reflecting the fact that all document updates are enacted through human actions.

This formulation highlights a key structural property of interactive co-writing:
while the assistant influences the process by proposing candidate continuations,
control over state transitions resides entirely with the human author.
As a result, learning and evaluation in this setting must be grounded
in observable human responses to system proposals.

\section{Co-Writing Fidelity Suite}

\begin{figure*}[t]
 \centering
  \includegraphics[width=1\linewidth]{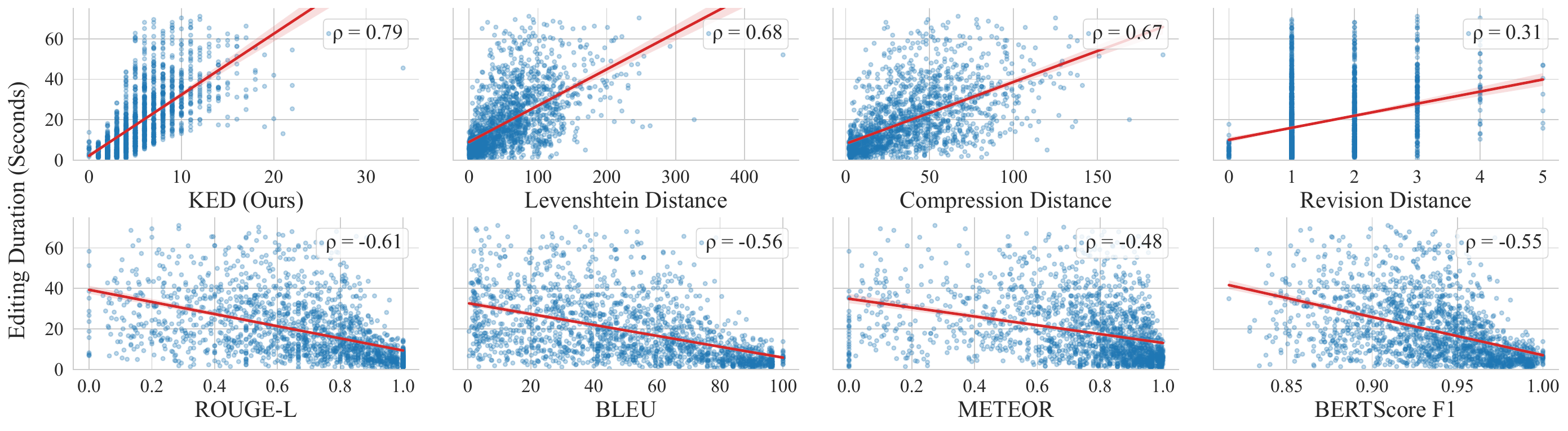}
  \caption {Editing duration versus editing distance and similarity metrics.
Each point represents a human post-editing instance; red lines show linear regression trends. KED aligns most strongly with observed editing time, whereas surface-level distances and similarity metrics correlate weakly with human editing effort.
}

\label{fig:cwed_scatter}
\end{figure*}

\label{sec:fidelity_suite}
Modeling mixed-initiative co-writing as a HitL-MDP makes user acceptance and editing actions explicit.
Yet these behaviors are costly, noisy, and ill-suited for large-scale or comparative evaluation across interaction paradigm.
To address this limitation, we adopt a \emph{human simulator} perspective, in which user feedback is abstracted into behavioral proxies grounded in observed co-writing patterns.
Under this view, a useful suggestion is not defined by fluency alone, but by whether it (i) satisfies the user’s implicit acceptance criteria and (ii) minimizes the cognitive effort required for integration.

Motivated by this perspective, we introduce the \textbf{Co-Writing Fidelity Suite}, a set of interaction-aware metrics that approximate human acceptance behavior and post-acceptance editing cost within the HitL-MDP framework.
Specifically, we propose \textbf{Hierarchical Acceptance Rate (HAR)} to measure intent-aligned acceptance, and \textbf{Knowledge-aware Editing Distance (KED)} to quantify the cognitive effort incurred after acceptance.

\begin{table}[t]
\centering
\small
\setlength{\tabcolsep}{2.8pt}  

\caption{Alignment rate (\%) and inter-annotator agreement (Cohen's $\kappa$). 
Best results are \textbf{bolded}; second-best are \underline{underlined}.}

\begin{tabular}{lcccccc}
\toprule
\multirow{2}{*}{\textbf{Metrics}} 
& \multicolumn{2}{c}{\textbf{DeepSeek-V3}} 
& \multicolumn{2}{c}{\textbf{GPT-5.1}} 
& \multicolumn{2}{c}{\textbf{Gemini-2.5-Pro}} \\
\cmidrule(lr){2-3}
\cmidrule(lr){4-5}
\cmidrule(lr){6-7}
& Align & $\kappa$
& Align & $\kappa$
& Align & $\kappa$ \\
\midrule

\rowcolor{gray!12}
\multicolumn{7}{l}{\textit{Baselines (Single-Dimension \& Holistic)}} \\

$\text{LLM}_{\text{Logic}}$ 
& 51.6 & 0.072
& 53.5 & 0.032
& 56.6 & 0.020 \\

$\text{LLM}_{\text{Style}}$  
& 67.3 & 0.106
& 51.4 & 0.102
& 57.7 & 0.018 \\

$\text{LLM}_{\text{Semantic}}$  
& \underline{69.0} & \underline{0.158}
& 74.6 & \textbf{0.184}
& 65.7 & 0.180 \\

$\text{LLM}_{\text{Holistic}}$  
& 48.8 & 0.014
& \underline{78.3} & 0.063
& 64.1 & 0.196 \\

\midrule
\rowcolor{gray!12}
\multicolumn{7}{l}{\textit{Hierarchical Ablation (Ours)}} \\

$\text{HAR}_{\text{L1}}$ 
& 59.9 & 0.126
& 76.3 & 0.045
& 67.0 & \textbf{0.269} \\

$\text{HAR}_{\text{L1+L2}}$
& 60.3 & 0.104
& 77.0 & 0.070
& \underline{79.8} & 0.243 \\

$\text{HAR}_{\text{Full}}$
& \textbf{69.3} & \textbf{0.166}
& \textbf{81.1} & \underline{0.154}
& \textbf{82.2} & \underline{0.254} \\

\bottomrule
\end{tabular}


\label{tab:HAR}
\end{table}

\begin{table}[t]
\centering
\small
\setlength{\tabcolsep}{2pt}  %

\caption{Correlation between automatic evaluation metrics and human post-editing effort,
measured by editing duration.
Pearson ($r$) and Spearman ($\rho$) correlations are reported.}

\begin{tabular}{lccc}
\toprule
\textbf{Metric} & \textbf{Pearson} & \textbf{Spearman}  \\
\midrule
\textbf{KED (ours)} & \textbf{0.686} & \textbf{0.786} \\
Levenshtein Distance  & 0.587 & 0.684  \\
Compression Distance   & 0.582 & 0.670  \\
Revision Distance   & 0.305 & 0.306 \\
ROUGE-L             & -0.498 & -0.612  \\
BLEU                & -0.499 & -0.565  \\
METEOR              & -0.356 & -0.482  \\
BERTScore           & -0.469 & -0.546  \\
\bottomrule
\end{tabular}

\label{tab:CWED}
\end{table}

\subsection{Hierarchical Acceptance Rate}
\paragraph{Motivation and Definition.}
In proactive co-writing, acceptance is a structured decision rather than a binary judgment.
Prior work in cognitive psychology and behavioral decision theory suggests that humans evaluate complex options through sequential filtering processes, progressively eliminating candidates that fail increasingly stringent criteria rather than making a single holistic judgment \cite{lee2024design,simon1955behavioral,tversky1972elimination,payne1993adaptive}.
Users thus implicitly apply a sequence of filters—ranging from basic correctness to higher-level semantic alignment—before integrating a suggestion.

We model this process with \emph{Hierarchical Decision Criteria}.
Given an ordered set of criterias $\{ \phi_1, \phi_2, \dots, \phi_K \}$, a suggestion is accepted only if it passes all criterias, where failure at any stage terminates the process.
Early-stage criteria capture immediate usability (e.g., grammaticality, continuation correctness), while later stages assess stylistic and semantic alignment, including required entity coverage.
When applied across a collection of model-generated suggestions, this hierarchical decision process induces a well-defined acceptance probability, which is formalized as \emph{Hierarchical Acceptance Rate}.
This conservative design ensures stable and interpretable acceptance judgments across interaction rounds.

\paragraph{Alignment with Human Evaluation.}
We evaluate HAR on 1.2K expert-annotated instances. Each instance contains a writing context, a model-generated continuation, and a gold reference continuation; annotators determine whether the suggestion should be accepted using a checklist-style protocol that mirrors realistic co-writing decisions.

We compare HAR against two representative LLM-as-a-Judge alternatives: \emph{single-dimension judges} that decide acceptance based only on logic, style, or semantic correctness, and a \emph{holistic judge} that outputs one-shot accept/reject without explicitly modeling intermediate criteria.
We report (i) \emph{alignment rate} with expert decisions and (ii) \emph{inter-annotator agreement} measured by Cohen’s $\kappa$.

\begin{table*}[t]
\centering
\small
\caption{Interaction paradigms organized by increasing levels of initiative, contextual awareness, and adaptivity.
The progression highlights the transition from reactive to proactive intent inference in human--AI co-writing.}

\label{tab:interaction_paradigm}

\begin{tabularx}{\textwidth}{c l l l}
\toprule
\textbf{Level} & \textbf{Interaction Paradigm} & \textbf{Interaction Scope} & \textbf{Key Capability} \\
\midrule
L0 & User-Initiated Collaboration
   & Reacting to explicitly stated user intent
   & Demand Responsiveness \\

L1 & Stateless Proactive Collaboration
   & Proactively predicting user intent from local context
   & Immediate Feedback \\

L2 & Stateful Proactive Collaboration 
   & Proactively predicting user intent from interaction history 
   & Experience-Based Personalization \\

L3 & Adaptive Proactive Collaboration 
   & Adaptively predicting user intent via strategies
   & Adaptive Intent Modeling \\
\bottomrule
\end{tabularx}

\end{table*}

Table~\ref{tab:HAR} shows that HAR yields the most reliable human-aligned acceptance estimates across backbone models.
In terms of alignment, the full hierarchical design (\textsc{HAR}$_\text{Full}$) achieves the best performance for all three backbones (e.g., 69.3/81.1/82.2 for DeepSeek-V3/GPT-5.1/Gemini-2.5-Pro), while single-dimension judges vary substantially across model families, indicating limited robustness when acceptance is reduced to a single criterion.

Agreement results further support the benefit of explicit criteria: \textsc{HAR}$_\text{Full}$ achieves the highest $\kappa$ on DeepSeek-V3 (0.166) and remains competitive on GPT-5.1 (0.154), and on Gemini-2.5-Pro the hierarchical variants markedly improve agreement (up to 0.269), suggesting that structuring acceptance into ordered criteria reduces ambiguity compared to one-shot holistic decisions.

Overall, these results are consistent with the view that acceptance in interactive co-writing is better characterized as a \emph{hierarchical decision process} rather than a monolithic judgment: making intermediate criteria explicit not only improves alignment with expert choices, but also produces more consistent acceptance decisions.

\subsection{Knowledge-aware Editing Distance}
\paragraph{Motivation and Definition.}
Acceptance alone does not imply low interaction cost.
In realistic co-writing, users accept a suggestion yet invest substantial effort in post-editing to reconcile it with their underlying intent.
Such effort is frequently driven by correcting semantic mismatches, factual inaccuracies, or discourse-level inconsistencies, which are poorly captured by surface-level edit-distance or similarity metrics.

We introduce \emph{Knowledge-aware Editing Distance} (KED) to quantify this post-acceptance editing cost.
KED measures the distance between a model-generated suggestion and the final user-edited text, while weighting edit operations according to their cognitive demands: semantic rewrites and factual corrections incur higher costs than local or surface-level edits.
As a result, KED is designed to reflect not just textual change, but the mental effort required to repair a suggestion into an acceptable continuation.

\paragraph{Alignment with Human Evaluation.}
We evaluate KED using writing logs from the CoAuthor dataset~\cite{lee2022coauthor}, which records human revisions to AI-generated suggestions together with timestamps.
Following prior work\cite{devatine2024assessing}, we use \emph{editing duration} as a behavioral proxy for cognitive effort, as longer interaction time is consistently associated with increased mental workload.
We compare KED against a diverse set of baselines, including character-level distances (Levenshtein Distance), structure-aware distances (Compression Distance\cite{devatine2024assessing}, Revision Distance\cite{ma2024RevisionDistance}), and similarity-based metrics (ROUGE-L, BLEU, METEOR, and BERTScore).
For each metric, we compute Pearson and Spearman correlations with observed editing duration.

Figure~\ref{fig:cwed_scatter} illustrates the relationship between editing duration and metrics.
KED exhibits the strongest monotonic and linear correlations with editing time, indicating that it more faithfully reflects the cognitive cost experienced by users during post-editing.
In contrast, conventional edit distances show weaker correlations, as uniform edit costs overemphasize superficial changes.
Similarity-based metrics display moderate to weak negative correlations, confirming that high textual similarity does not necessarily imply low editing effort.

Table~\ref{tab:CWED} quantifies these trends.
KED achieves the highest Pearson and Spearman coefficients among all evaluated metrics, demonstrating both sensitivity to absolute editing effort and consistency in ranking instances by difficulty.
All reported correlations are statistically significant.

Taken together, our findings indicate that model suggestions vary in quality even under universal acceptance.
Explicitly modeling the cognitive burden of different edit types is essential for capturing the friction users experience after accepting an assistant-generated suggestion.

\section{Co-Writing Experiment}
\label{sec:experiments}

\begin{table}[t]
\centering
\setlength{\tabcolsep}{2pt} 

\caption{Results of LLMs under Level-1 and Level-2 interaction paradigm across scientific and creative domain.}

\resizebox{\columnwidth}{!}{
\begin{tabular}{lllcccccc}
\toprule
\multirow{2}{*}{\textbf{Model}} &
\multirow{2}{*}{\textbf{Level}} 

& \multicolumn{2}{c}{\textbf{Overall}} 
& \multicolumn{2}{c}{\textbf{Scientific}}
& \multicolumn{2}{c}{\textbf{Creative}} \\

\cmidrule(lr){3-4}
\cmidrule(lr){5-6}
\cmidrule(lr){7-8}
& & HAR(\%) $\uparrow$ & KED $\downarrow$
& HAR(\%) $\uparrow$ & KED $\downarrow$
& HAR(\%) $\uparrow$ & KED $\downarrow$ \\

\midrule
\multirow{2}{*}{DeepSeek-V3.2} & 
L1 & 6.99 & 10.18  & 8.18 & 11.70 & 5.35 & 7.11  \\
& L2 & 10.37 & 9.97  & 12.88 & 11.39 & 6.90 & 6.31 \\

\midrule

\multirow{2}{*}{Gemini-2.5-Flash} & 
L1 & 13.39 & 9.68 & 14.83 & 11.11 & 11.41 & 7.20  \\
& L2 & 16.05 & 10.24 & 20.04 & 11.79 & 10.56 & 6.35  \\

\midrule

\multirow{2}{*}{Gemini-2.5-Pro} & 
L1 & 14.17 & \textbf{9.17}  & 15.15 & 11.00 & 12.82 & 6.20  \\
& L2 & 17.89 & 9.39 & 19.12 & \textbf{10.81} & \textbf{16.20} & 7.03  \\

\midrule

\multirow{2}{*}{GPT-5.1} & 
L1 & 17.59 & 11.28  & 21.47 & 12.05 & 12.25 & 9.46  \\
& L2 & \textbf{19.61} & 10.50 &  \textbf{23.62} & 12.14 & 14.08 & \textbf{6.06} \\

\bottomrule
\end{tabular}
}

\label{tab:main_results}
\end{table}

\subsection{Interaction Paradigms}
\label{sec:interaction_paradigms}

We organize co-writing systems into four interaction paradigms,
as summarized in Table~\ref{tab:interaction_paradigm}.
These paradigms differ in how initiative is distributed between the user and the assistant,
what contextual signals are available for intent inference,
and whether the system adapts its intervention behavior over time.
Together, they characterize a structured progression
from passive assistance to proactive and adaptive collaboration,
providing a unified lens for analyzing user acceptance and editing behavior.

\paragraph{L0: User-Initiated Collaboration.}
In this paradigm, the assistant responds only to explicitly stated user intent
and provides assistance on demand.
All initiative remains with the user, and the system does not anticipate or infer intent.
L0 emphasizes demand responsiveness and serves as a reference condition,
particularly in user-facing evaluations.

\paragraph{L1: Stateless Proactive Collaboration.}
L1 introduces system-initiated assistance by proactively predicting
immediate user intent from local document context.
Suggestions are generated independently at predefined points,
without access to prior interaction outcomes.
This paradigm provides immediate feedback while isolating the effect of proactive intervention.

\paragraph{L2: Stateful Proactive Collaboration.}
L2 extends L1 by conditioning intent prediction on accumulated interaction history,
including previous human feedback.
This enables experience-based personalization,
allowing the assistant to progressively align with the user's evolving intent.
Comparing L2 with L1 isolates the behavioral impact of interaction memory.

\paragraph{L3: Adaptive Proactive Collaboration.}
L3 further introduces adaptive intent modeling,
where the assistant dynamically determines when and how to provide suggestions.
Intervention itself becomes a strategy-level decision,
making interaction events endogenous to the system policy.
We therefore treat L3 as an analysis-driven paradigm,
informed by insights from L1 and L2,
and reserve its evaluation for downstream system design and user studies.

In the following sections, we focus quantitative analysis on L1 and L2,
where interaction events are well-defined and directly comparable,
and use the resulting insights to inform adaptive collaboration in L3.

\subsection{Quantitative Evaluation}

\paragraph{Setup.}
We evaluate multiple language models under different interaction paradigms using a controlled dataset of 60 human-authored articles across 16 domains. Articles are segmented into 1,688 continuation queries sampled from different writing stages, with fixed ground-truth continuations across settings to isolate the effect of interaction structure.

\begin{figure}[t]
  \includegraphics[width=0.99\linewidth]{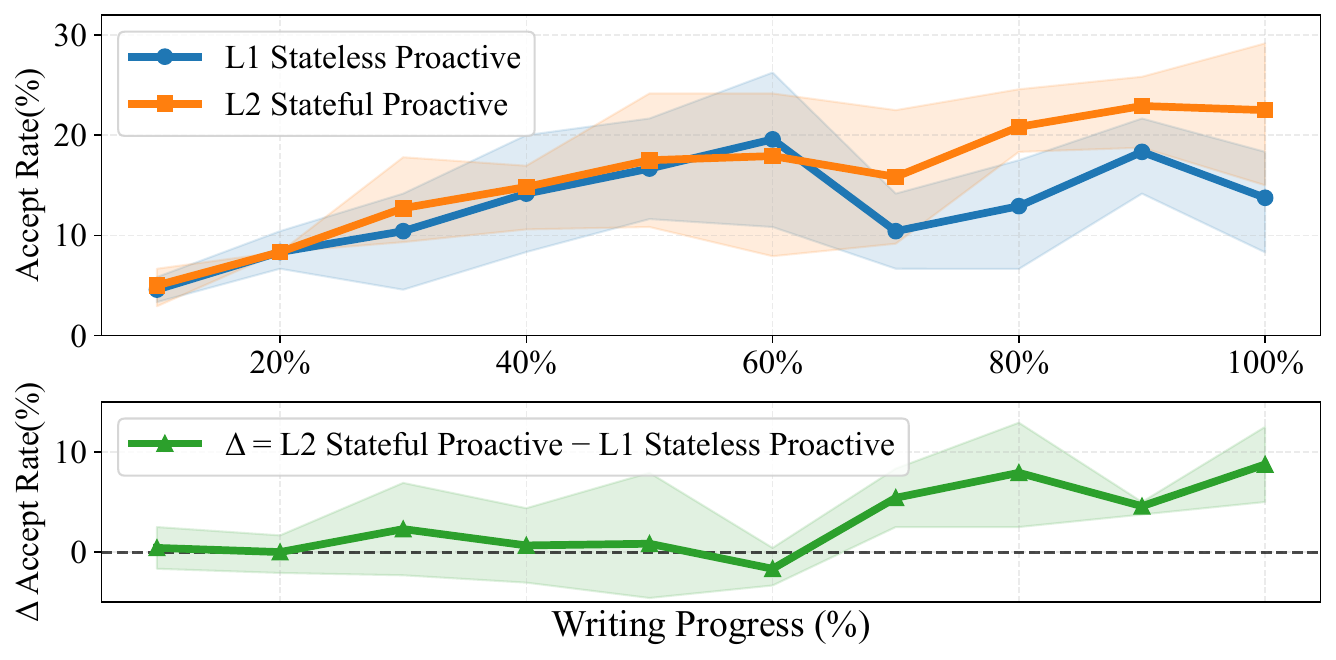}
  \caption {
Acceptance rate as a function of writing progress for L1 and L2 proactive interaction. L2 exhibits increasing advantages in later stages, with $\Delta$ curves and confidence intervals shown below.
}
\label{fig:multi_vs_single}
\end{figure}

\begin{figure}[t]
  \includegraphics[width=0.99\linewidth]{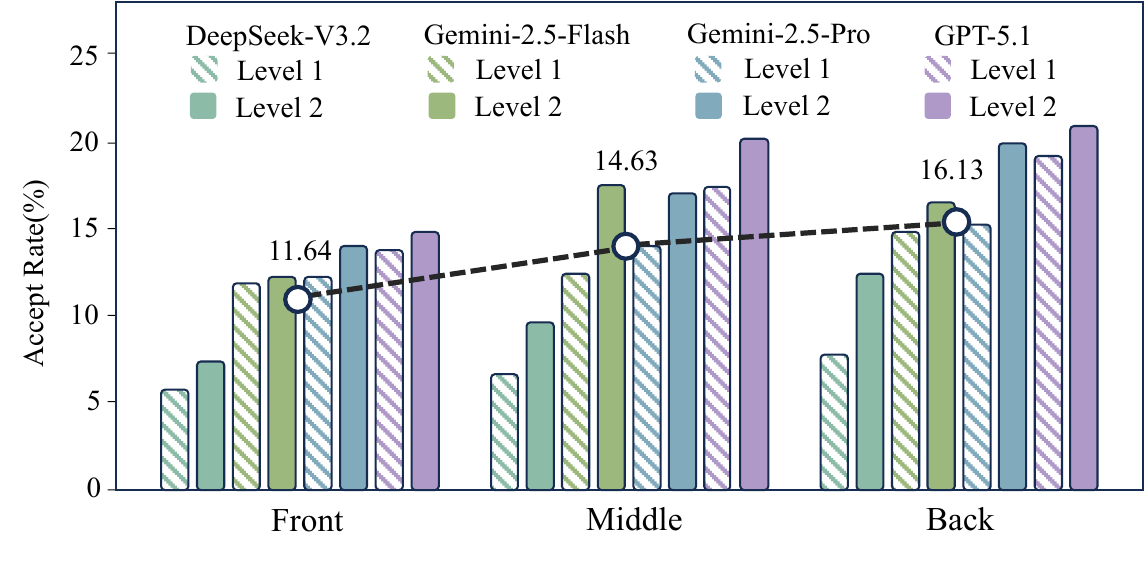}
  \caption {Paragraph-level acceptance rates increase from early to late paragraphs across models and interaction settings.
}
\label{fig:stage}
\end{figure}

\paragraph{Results and Analysis.}
As shown in Table~5, GPT-5.1 achieves the highest HAR overall, while Gemini-2.5-Pro attains the lowest KED, reflecting complementary strengths in intent prediction and post-editing efficiency. Moving from L1 to L2 further yields systematic gains in acceptance and reduced editing effort, underscoring the value of stateful intent modeling.
\begin{itemize}
    \item \textbf{Intent prediction accuracy increases as writing progresses.}
    As shown in Figure\ref{fig:multi_vs_single}, both stateless (L1) and stateful (L2) proactive settings exhibit an upward trend in acceptance rate from early to later writing stages, indicating that richer local context facilitates more accurate intent inference. Notably, the performance gap between L2 and L1 widens in the later stages of writing, where accumulated interaction history becomes available. The consistently positive $\Delta$ curve, together with its confidence intervals, suggests that conditioning on prior user decisions enables more effective anticipation of user intent beyond what can be inferred from local context alone. These results highlight the growing value of interaction memory as co-writing proceeds and user intent becomes increasingly shaped by earlier choices.
    
    \item \textbf{Paragraph-level acceptance increases monotonically with writing progress.} As shown in Figure~\ref{fig:stage}, acceptance rates computed from paragraph-level continuations consistently rise from the front to the middle and back sections of an article across all evaluated models and interaction settings. This monotonic trend holds for both stateless and stateful proactive paradigms, as well as across different model families, indicating that the effect is not model-specific. The increasing acceptance toward later paragraphs suggests that author intent becomes progressively more explicit and constrained at the paragraph level, making intent inference more reliable as the document structure and thematic direction solidify.
    
    \item \textbf{Acceptance rates are consistently lower in Creative than Scientific domains.} This pattern is consistently observed across models and interaction settings (Table\ref{tab:main_results}) and is characteristic of the higher semantic openness of creative writing.
\end{itemize}

Based on the above analyses, L3 Collaboration should treat proactivity as a cost-sensitive policy conditioned on interaction history, writing stage, and domain characteristics. Such adaptivity aligns system initiative with the evolving predictability of user intent while limiting unnecessary editing overhead.

\subsection{User Study}
To validate whether the behavioral differences observed in offline evaluation
translate into real user experience,
we conduct a controlled user study with 30 participants recruited from universities and technology companies, using an online interactive writing system
that instantiates all four interaction paradigms with a unified LLM backend (GPT-5.1).
User experience is assessed using Likert-scale measures of perceived workload,
usefulness, customization, and adaptability,
following established human--AI interaction guidelines~\cite{amershi2019guidelines,google_pair_hai_2021,apple_hig_ml_2023,pasch2025human}, and supplemented with qualitative feedback.

\begin{figure}[t]
 \centering
  \includegraphics[width=1\linewidth]{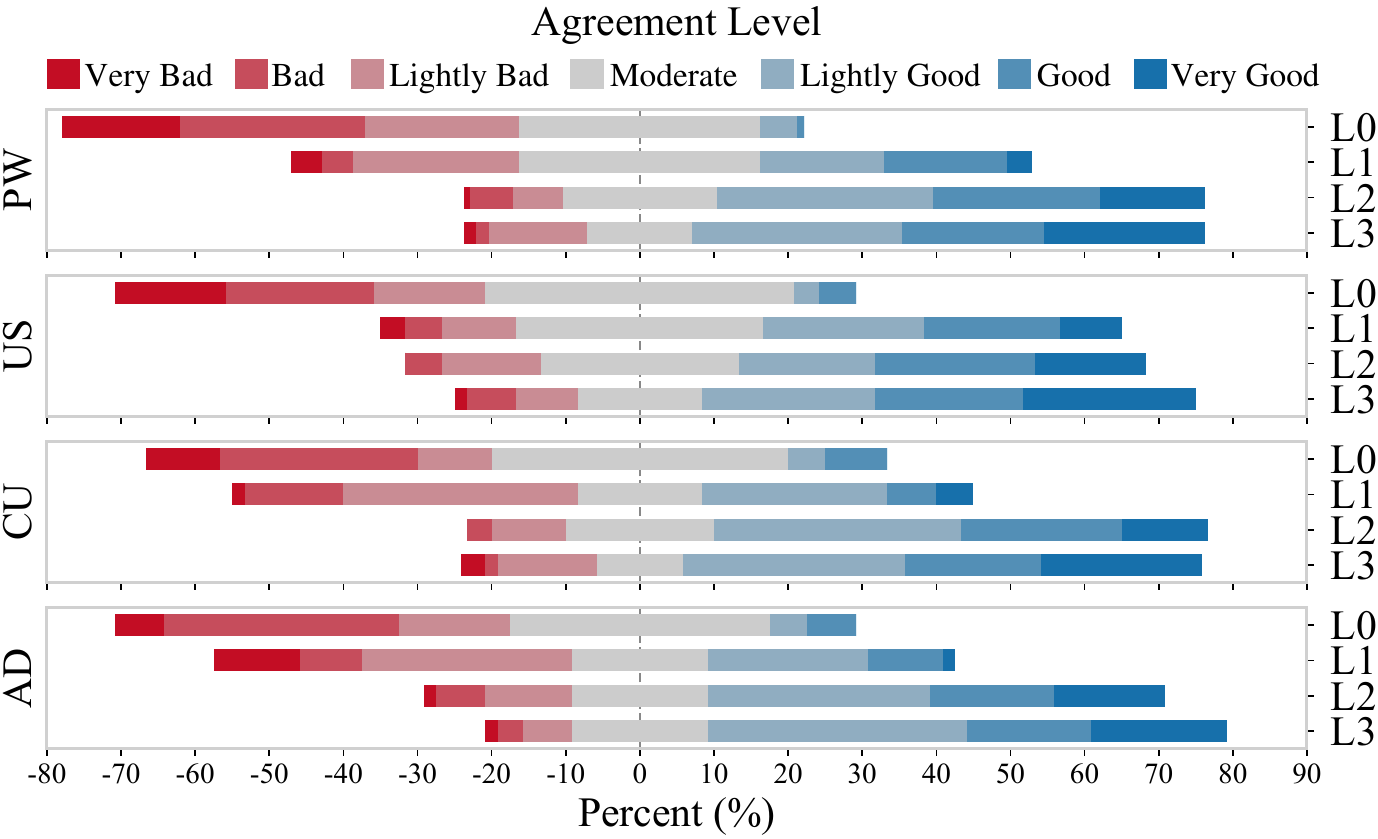}
  \caption {User questionnaire results across four interaction systems for perceived workload(PW), usefulness(US), customization(CU), and adaptability(AD). Blue indicates more favorable ratings, while red indicates less favorable ratings.
}
\label{fig:userstudy1}
\end{figure}

As shown in Figure~\ref{fig:userstudy1} and Table~\ref{tab:user_study_log},
both subjective ratings and interaction logs exhibit consistent trends.
Proactive paradigms (L1--L3) are preferred over user-initiated interaction (L0),
with stateful and adaptive settings (L2, L3) achieving the lowest perceived workload,
highest usefulness, and strongest adaptability.
Interaction logs further corroborate these perceptions:
proactive collaboration substantially reduces window switching
and task completion time, while acceptance rates increase
from stateless to stateful and adaptive settings.
Together, these results confirm that the gains identified in quantitative analysis
reflect genuine improvements in user experience rather than artifacts of offline metrics.

\begin{table}[t]
\centering
\setlength{\tabcolsep}{3pt}  

\caption{User study results based on interaction logs, reporting mean window switch counts, editing time, and acceptance rate averaged across all users under different interaction levels.}

\resizebox{\columnwidth}{!}{
\begin{tabular}{lccc}
\toprule
\textbf{Level} & \textbf{Window Switches $\downarrow$}(count) & \textbf{Editing Time $\downarrow$} (min) & \textbf{Acceptance Rate $\uparrow$} (\%)  \\
\midrule
L0
  & 15.76
  & 46.42
  & - 
  \\
L1
  & 8.46
  & 32.85
  & 18.55 \\
L2
  & 8.92
  & \textbf{31.07}
  & 21.62
  \\
L3
  & \textbf{8.15}
  & 31.28
  & \textbf{27.87} \\
\bottomrule
\end{tabular}
}

\label{tab:user_study_log}
\end{table}

\begin{figure}[t]
 \centering
  \includegraphics[width=1\linewidth]{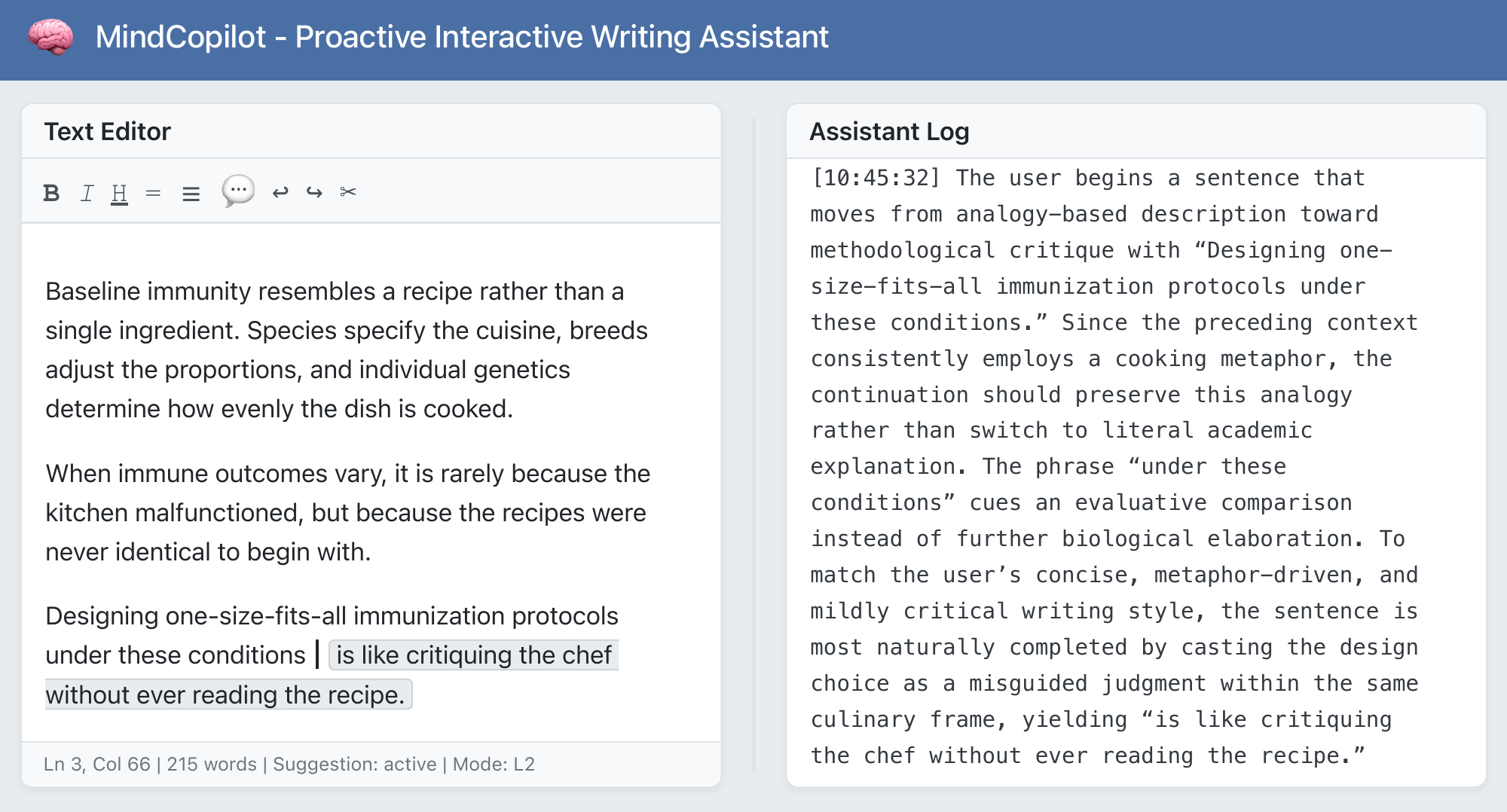}
  
  \caption {A stateful co-writing interface in which the assistant leverages interaction history to infer the user’s preference for analogy-driven critique and predicts a style-consistent continuation.
}
\label{fig:demoUI}
\end{figure}

\begin{table}[t]
\centering
\small
\caption{Performance of Qwen3-4B models after Stateless RL and Stateful RL. L1 and L2 Collaboration are defined in Table \ref{tab:interaction_paradigm}.}
\label{tab:qwen3_behavior_rl}

\resizebox{\columnwidth}{!}{
\begin{tabular}{lcccc}
\toprule
\multirow{2}{*}{\textbf{Model}}
& \multicolumn{2}{c}{\textbf{L1 Collaboration}}
& \multicolumn{2}{c}{\textbf{L2 Collaboration}}
\\

\cmidrule(lr){2-3}
\cmidrule(lr){4-5}
 & HAR(\%) $\uparrow$ & KED $\downarrow$ & HAR(\%) $\uparrow$ & KED $\downarrow$
\\

\midrule
Qwen3-4B 
    & 1.48  & 14.79 & 0.95 & 15.66 \\

\quad + Stateless RL
    & 8.30 & 12.64 & 7.47 & 12.38 \\

\quad + Stateful RL
    & \textbf{8.86} & \textbf{12.56} & \textbf{9.73} & \textbf{12.28} \\
\bottomrule
\end{tabular}
}

\end{table}

\subsection{Learning with Interaction-Aware Rewards}
Finally, we conduct reinforcement learning experiments to examine the practical learnability of our interaction-aware metric under the proposed formalization.
We conduct reinforcement learning experiments on Qwen3-4B~\cite{yang2025qwen3} using HAR as the sole reward for continuation prediction, considering two training variants. In \emph{stateless RL}, the model is optimized using the original query context, without access to prior interaction history. In \emph{stateful RL}, interaction history is incorporated into the context during training.
As shown in Table~\ref{tab:qwen3_behavior_rl}, both RL variants substantially improve HAR and reduce KED compared to the base model.
Moreover, stateful RL consistently outperforms stateless RL, particularly under L2 collaboration, indicating that modeling interaction history further enhances learning. These results confirm that HAR can be effectively used as a reward for training, providing empirical support that our Human-in-the-Loop MDP formulation admits practical optimization and yields measurable gains in interactive writing performance.
\section{Discussions and Conclusion}

This work advances the study of human–AI co-writing by shifting the focus from isolated output quality to interaction-driven user behavior. By formalizing interactive completion as a Human-in-the-Loop Markov Decision Process, we provide a framework for analyzing how acceptance and editing actions reveal alignment and friction during writing.

The proposed Co-Writing Fidelity Suite evaluates collaboration through behavioral signals. \emph{Hierarchical Acceptance Rate} captures the structured nature of human acceptance decisions, while \emph{Knowledge-aware Editing Distance} quantifies post-acceptance cognitive effort. Across expert annotation, large-scale simulation, and a controlled user study, these metrics demonstrate stronger alignment with human judgment and user experience than output-only measures.

Using these metrics as analytical tools, we identify consistent behavioral patterns across interaction settings and writing stages, and derive interaction strategies that better match users’ evolving intent. Importantly, systems informed by these metric-driven analyses yield higher perceived usefulness and lower cognitive burden in real writing tasks, supporting the external validity of the proposed evaluation approach.

\section*{Acknowledgements}
We thank the anonymous reviewers and area chair for their helpful comments.

\bibliographystyle{named}
\bibliography{ijcai26}

\begin{thebibliography}{}

\bibitem[\protect\citeauthoryear{Amershi \bgroup \em et al.\egroup }{2019}]{amershi2019guidelines}
Saleema Amershi, Dan Weld, Mihaela Vorvoreanu, Adam Fourney, Besmira Nushi, Penny Collisson, Jina Suh, Shamsi Iqbal, Paul~N Bennett, Kori Inkpen, et~al.
\newblock Guidelines for human-ai interaction.
\newblock In {\em Proceedings of the 2019 chi conference on human factors in computing systems}, pages 1--13, 2019.

\bibitem[\protect\citeauthoryear{{Anthropic}}{2025}]{anthropic_us_usage}
{Anthropic}.
\newblock Economic index -- us usage, 2025.

\bibitem[\protect\citeauthoryear{{Apple}}{2023}]{apple_hig_ml_2023}
{Apple}.
\newblock {Human Interface Guidelines: Machine Learning}.
\newblock \url{https://developer.apple.com/design/human-interface-guidelines/machine-learning}, 2023.
\newblock Accessed: 2026-01-13.

\bibitem[\protect\citeauthoryear{Ashkinaze \bgroup \em et al.\egroup }{2025}]{ashkinaze2025ai}
Joshua Ashkinaze, Julia Mendelsohn, Li~Qiwei, Ceren Budak, and Eric Gilbert.
\newblock How ai ideas affect the creativity, diversity, and evolution of human ideas: evidence from a large, dynamic experiment.
\newblock In {\em Proceedings of the ACM collective intelligence conference}, pages 198--213, 2025.

\bibitem[\protect\citeauthoryear{Bai \bgroup \em et al.\egroup }{2022}]{bai2022training}
Yuntao Bai, Andy Jones, Kamal Ndousse, Amanda Askell, Anna Chen, Nova DasSarma, Dawn Drain, Stanislav Fort, Deep Ganguli, Tom Henighan, et~al.
\newblock Training a helpful and harmless assistant with reinforcement learning from human feedback.
\newblock {\em arXiv preprint arXiv:2204.05862}, 2022.

\bibitem[\protect\citeauthoryear{Bai \bgroup \em et al.\egroup }{2024}]{bai2408longwriter}
Yushi Bai, Jiajie Zhang, Xin Lv, Linzhi Zheng, Siqi Zhu, Lei Hou, Yuxiao Dong, Jie Tang, and Juanzi Li.
\newblock Longwriter: Unleashing 10,000+ word generation from long context llms, 2024.
\newblock {\em URL https://arxiv. org/abs/2408.07055}, 2024.

\bibitem[\protect\citeauthoryear{Chung \bgroup \em et al.\egroup }{2025}]{chung2025literarytaste}
John Joon~Young Chung, Vishakh Padmakumar, Melissa Roemmele, Yi~Wang, Yuqian Sun, Tiffany Wang, Shm~Garanganao Almeda, Brett~A Halperin, Yuwen Lu, and Max Kreminski.
\newblock Literarytaste: A preference dataset for creative writing personalization.
\newblock {\em arXiv preprint arXiv:2511.09310}, 2025.

\bibitem[\protect\citeauthoryear{Coenen \bgroup \em et al.\egroup }{2021}]{coenen2021wordcraft}
Andy Coenen, Luke Davis, Daphne Ippolito, Emily Reif, and Ann Yuan.
\newblock Wordcraft: a human-ai collaborative editor for story writing.
\newblock {\em arXiv preprint arXiv:2107.07430}, 2021.

\bibitem[\protect\citeauthoryear{Comanici \bgroup \em et al.\egroup }{2025}]{comanici2025gemini}
Gheorghe Comanici, Eric Bieber, Mike Schaekermann, Ice Pasupat, Noveen Sachdeva, Inderjit Dhillon, Marcel Blistein, Ori Ram, Dan Zhang, Evan Rosen, et~al.
\newblock Gemini 2.5: Pushing the frontier with advanced reasoning, multimodality, long context, and next generation agentic capabilities.
\newblock {\em arXiv preprint arXiv:2507.06261}, 2025.

\bibitem[\protect\citeauthoryear{Devatine and Abraham}{2024}]{devatine2024assessing}
Nicolas Devatine and Louis Abraham.
\newblock Assessing human editing effort on llm-generated texts via compression-based edit distance.
\newblock {\em arXiv preprint arXiv:2412.17321}, 2024.

\bibitem[\protect\citeauthoryear{Dhillon \bgroup \em et al.\egroup }{2024}]{dhillon2024shaping}
Paramveer~S Dhillon, Somayeh Molaei, Jiaqi Li, Maximilian Golub, Shaochun Zheng, and Lionel~Peter Robert.
\newblock Shaping human-ai collaboration: Varied scaffolding levels in co-writing with language models.
\newblock In {\em Proceedings of the 2024 CHI Conference on Human Factors in Computing Systems}, pages 1--18, 2024.

\bibitem[\protect\citeauthoryear{Du \bgroup \em et al.\egroup }{2025}]{du2025deepresearch}
Mingxuan Du, Benfeng Xu, Chiwei Zhu, Xiaorui Wang, and Zhendong Mao.
\newblock Deepresearch bench: A comprehensive benchmark for deep research agents.
\newblock {\em arXiv preprint arXiv:2506.11763}, 2025.

\bibitem[\protect\citeauthoryear{Fang \bgroup \em et al.\egroup }{2025}]{fang2025creation}
Xinyu Fang, Zhijian Chen, Kai Lan, Lixin Ma, Shengyuan Ding, Yingji Liang, Xiangyu Zhao, Farong Wen, Zicheng Zhang, Guofeng Zhang, et~al.
\newblock Creation-mmbench: Assessing context-aware creative intelligence in mllm.
\newblock {\em arXiv preprint arXiv:2503.14478}, 2025.

\bibitem[\protect\citeauthoryear{Fein \bgroup \em et al.\egroup }{2025}]{fein2025litbench}
Daniel Fein, Sebastian Russo, Violet Xiang, Kabir Jolly, Rafael Rafailov, and Nick Haber.
\newblock Litbench: A benchmark and dataset for reliable evaluation of creative writing.
\newblock {\em arXiv preprint arXiv:2507.00769}, 2025.

\bibitem[\protect\citeauthoryear{{Google}}{2021}]{google_pair_hai_2021}
{Google}.
\newblock {People + AI Design Guidebook}.
\newblock \url{https://pair.withgoogle.com/guidebook/patterns}, 2021.
\newblock Accessed: 2026-01-13.

\bibitem[\protect\citeauthoryear{He \bgroup \em et al.\egroup }{2025}]{he2025contributions}
Jessica He, Stephanie Houde, and Justin~D Weisz.
\newblock Which contributions deserve credit? perceptions of attribution in human-ai co-creation.
\newblock In {\em Proceedings of the 2025 CHI Conference on Human Factors in Computing Systems}, pages 1--18, 2025.

\bibitem[\protect\citeauthoryear{Jia \bgroup \em et al.\egroup }{2025}]{jia2025writing}
Ruipeng Jia, Yunyi Yang, Yongbo Gai, Kai Luo, Shihao Huang, Jianhe Lin, Xiaoxi Jiang, and Guanjun Jiang.
\newblock Writing-zero: Bridge the gap between non-verifiable tasks and verifiable rewards.
\newblock {\em arXiv e-prints}, pages arXiv--2506, 2025.

\bibitem[\protect\citeauthoryear{Laban \bgroup \em et al.\egroup }{2024}]{laban2024beyond}
Philippe Laban, Jesse Vig, Marti Hearst, Caiming Xiong, and Chien-Sheng Wu.
\newblock Beyond the chat: Executable and verifiable text-editing with llms.
\newblock In {\em Proceedings of the 37th Annual ACM Symposium on User Interface Software and Technology}, pages 1--23, 2024.

\bibitem[\protect\citeauthoryear{Le \bgroup \em et al.\egroup }{2025}]{le2025scholawrite}
Khanh~Chi Le, Linghe Wang, Minhwa Lee, Ross Volkov, Luan~Tuyen Chau, and Dongyeop Kang.
\newblock Scholawrite: A dataset of end-to-end scholarly writing process.
\newblock {\em arXiv preprint arXiv:2502.02904}, 2025.

\bibitem[\protect\citeauthoryear{Lee \bgroup \em et al.\egroup }{2022}]{lee2022coauthor}
Mina Lee, Percy Liang, and Qian Yang.
\newblock Coauthor: Designing a human-ai collaborative writing dataset for exploring language model capabilities.
\newblock In {\em Proceedings of the 2022 CHI conference on human factors in computing systems}, pages 1--19, 2022.

\bibitem[\protect\citeauthoryear{Lee \bgroup \em et al.\egroup }{2024}]{lee2024design}
Mina Lee, Katy~Ilonka Gero, John Joon~Young Chung, Simon~Buckingham Shum, Vipul Raheja, Hua Shen, Subhashini Venugopalan, Thiemo Wambsganss, David Zhou, Emad~A Alghamdi, et~al.
\newblock A design space for intelligent and interactive writing assistants.
\newblock In {\em Proceedings of the 2024 CHI Conference on Human Factors in Computing Systems}, pages 1--35, 2024.

\bibitem[\protect\citeauthoryear{Li \bgroup \em et al.\egroup }{2025}]{li2025coig}
Yunwen Li, Shuangshuang Ying, Xingwei Qu, Xin Li, Sheng Jin, Minghao Liu, Zhoufutu Wen, Tianyu Zheng, Xeron Du, Qiguang Chen, et~al.
\newblock Coig-writer: A high-quality dataset for chinese creative writing with thought processes.
\newblock {\em arXiv preprint arXiv:2510.14763}, 2025.

\bibitem[\protect\citeauthoryear{Li \bgroup \em et al.\egroup }{2026}]{li2026state}
Pengcheng Li, Jie Zhang, Tianwei Zhang, Han Qiu, Weiming Zhang, Nenghai Yu, Wenbo Zhou, et~al.
\newblock State-dependent safety failures in multi-turn language model interaction.
\newblock {\em arXiv preprint arXiv:2603.15684}, 2026.

\bibitem[\protect\citeauthoryear{Liao \bgroup \em et al.\egroup }{2025}]{liao2025rlmr}
Jianxing Liao, Tian Zhang, Xiao Feng, Yusong Zhang, Rui Yang, Haorui Wang, Bosi Wen, Ziying Wang, and Runzhi Shi.
\newblock Rlmr: Reinforcement learning with mixed rewards for creative writing.
\newblock {\em arXiv preprint arXiv:2508.18642}, 2025.

\bibitem[\protect\citeauthoryear{Liu \bgroup \em et al.\egroup }{2025}]{liu2025deepseek}
Aixin Liu, Aoxue Mei, Bangcai Lin, Bing Xue, Bingxuan Wang, Bingzheng Xu, Bochao Wu, Bowei Zhang, Chaofan Lin, Chen Dong, et~al.
\newblock Deepseek-v3. 2: Pushing the frontier of open large language models.
\newblock {\em arXiv preprint arXiv:2512.02556}, 2025.

\bibitem[\protect\citeauthoryear{Ma \bgroup \em et al.\egroup }{2024}]{ma2024RevisionDistance}
Yongqiang Ma, Lizhi Qing, Jiawei Liu, Yangyang Kang, Yue Zhang, Wei Lu, Xiaozhong Liu, and Qikai Cheng.
\newblock From model-centered to human-centered: Revision distance as a metric for text evaluation in llms-based applications.
\newblock {\em arXiv preprint arXiv:2404.07108}, 2024.

\bibitem[\protect\citeauthoryear{Mirowski \bgroup \em et al.\egroup }{2023}]{mirowski2023co}
Piotr Mirowski, Kory~W Mathewson, Jaylen Pittman, and Richard Evans.
\newblock Co-writing screenplays and theatre scripts with language models: Evaluation by industry professionals.
\newblock In {\em Proceedings of the 2023 CHI conference on human factors in computing systems}, pages 1--34, 2023.

\bibitem[\protect\citeauthoryear{Mysore \bgroup \em et al.\egroup }{2025}]{mysore2025prototypical}
Sheshera Mysore, Debarati Das, Hancheng Cao, and Bahareh Sarrafzadeh.
\newblock Prototypical human-ai collaboration behaviors from llm-assisted writing in the wild.
\newblock {\em arXiv preprint arXiv:2505.16023}, 2025.

\bibitem[\protect\citeauthoryear{{OpenAI}}{2025}]{openai_how_people_using}
{OpenAI}.
\newblock How people are using chatgpt, 2025.

\bibitem[\protect\citeauthoryear{Pasch and Ha}{2025}]{pasch2025human}
Stefan Pasch and Sun-Young Ha.
\newblock Human--ai interaction and user satisfaction: Empirical evidence from online reviews of ai products.
\newblock {\em International Journal of Human--Computer Interaction}, pages 1--22, 2025.

\bibitem[\protect\citeauthoryear{Payne \bgroup \em et al.\egroup }{1993}]{payne1993adaptive}
John~W Payne, James~R Bettman, and Eric~J Johnson.
\newblock {\em The adaptive decision maker}.
\newblock Cambridge university press, 1993.

\bibitem[\protect\citeauthoryear{Que \bgroup \em et al.\egroup }{2024}]{que2024hellobench}
Haoran Que, Feiyu Duan, Liqun He, Yutao Mou, Wangchunshu Zhou, Jiaheng Liu, Wenge Rong, Zekun~Moore Wang, Jian Yang, Ge~Zhang, et~al.
\newblock Hellobench: Evaluating long text generation capabilities of large language models.
\newblock {\em arXiv preprint arXiv:2409.16191}, 2024.

\bibitem[\protect\citeauthoryear{Reza \bgroup \em et al.\egroup }{2025}]{reza2025co}
Mohi Reza, Jeb Thomas-Mitchell, Peter Dushniku, Nathan Laundry, Joseph~Jay Williams, and Anastasia Kuzminykh.
\newblock Co-writing with ai, on human terms: Aligning research with user demands across the writing process.
\newblock {\em Proceedings of the ACM on Human-Computer Interaction}, 9(7):1--37, 2025.

\bibitem[\protect\citeauthoryear{Shen and Wu}{2023}]{shen2023parachute}
Hua Shen and Tongshuang Wu.
\newblock Parachute: Evaluating interactive human-lm co-writing systems.
\newblock {\em arXiv preprint arXiv:2303.06333}, 2023.

\bibitem[\protect\citeauthoryear{Simon}{1955}]{simon1955behavioral}
Herbert~A Simon.
\newblock A behavioral model of rational choice.
\newblock {\em The quarterly journal of economics}, pages 99--118, 1955.

\bibitem[\protect\citeauthoryear{Singh \bgroup \em et al.\egroup }{2025}]{singh2025openai}
Aaditya Singh, Adam Fry, Adam Perelman, Adam Tart, Adi Ganesh, Ahmed El-Kishky, Aidan McLaughlin, Aiden Low, AJ~Ostrow, Akhila Ananthram, et~al.
\newblock Openai gpt-5 system card.
\newblock {\em arXiv preprint arXiv:2601.03267}, 2025.

\bibitem[\protect\citeauthoryear{Tversky}{1972}]{tversky1972elimination}
Amos Tversky.
\newblock Elimination by aspects: A theory of choice.
\newblock {\em Psychological review}, 79(4):281, 1972.

\bibitem[\protect\citeauthoryear{Wu \bgroup \em et al.\egroup }{2025}]{wu2503writingbench}
Yuning Wu, Jiahao Mei, Ming Yan, Chenliang Li, Shaopeng Lai, Yuran Ren, Zijia Wang, Ji~Zhang, Mengyue Wu, Qin Jin, et~al.
\newblock Writingbench: A comprehensive benchmark for generative writing, march 2025.
\newblock {\em arXiv preprint arXiv:2503.05244}, 2025.

\bibitem[\protect\citeauthoryear{Yang \bgroup \em et al.\egroup }{2022a}]{yang2022ai}
Daijin Yang, Yanpeng Zhou, Zhiyuan Zhang, Toby Jia-Jun Li, and Ray Lc.
\newblock Ai as an active writer: Interaction strategies with generated text in human-ai collaborative fiction writing.
\newblock In {\em Joint proceedings of the ACM IUI workshops}, volume~10, pages 1--11. CEUR-WS Team, 2022.

\bibitem[\protect\citeauthoryear{Yang \bgroup \em et al.\egroup }{2022b}]{yang2022re3}
Kevin Yang, Yuandong Tian, Nanyun Peng, and Dan Klein.
\newblock Re3: Generating longer stories with recursive reprompting and revision.
\newblock {\em arXiv preprint arXiv:2210.06774}, 2022.

\bibitem[\protect\citeauthoryear{Yang \bgroup \em et al.\egroup }{2023}]{yang2023doc}
Kevin Yang, Dan Klein, Nanyun Peng, and Yuandong Tian.
\newblock Doc: Improving long story coherence with detailed outline control.
\newblock In {\em Proceedings of the 61st Annual Meeting of the Association for Computational Linguistics (Volume 1: Long Papers)}, pages 3378--3465, 2023.

\bibitem[\protect\citeauthoryear{Yang \bgroup \em et al.\egroup }{2025}]{yang2025qwen3}
An~Yang, Anfeng Li, Baosong Yang, Beichen Zhang, Binyuan Hui, Bo~Zheng, Bowen Yu, Chang Gao, Chengen Huang, Chenxu Lv, et~al.
\newblock Qwen3 technical report.
\newblock {\em arXiv preprint arXiv:2505.09388}, 2025.

\bibitem[\protect\citeauthoryear{Ying \bgroup \em et al.\egroup }{2025}]{ying2025beyond}
Shuangshuang Ying, Yunwen Li, Xingwei Qu, Xin Li, Sheng Jin, Minghao Liu, Zhoufutu Wen, Xeron Du, Tianyu Zheng, Yichi Zhang, et~al.
\newblock Beyond correctness: Evaluating subjective writing preferences across cultures.
\newblock {\em arXiv preprint arXiv:2510.14616}, 2025.

\bibitem[\protect\citeauthoryear{Zhang \bgroup \em et al.\egroup }{2025}]{zhang2025exploring}
Shuning Zhang, Hui Wang, and Xin Yi.
\newblock Exploring collaboration patterns and strategies in human-ai co-creation through the lens of agency: A scoping review of the top-tier hci literature.
\newblock {\em Proceedings of the ACM on Human-Computer Interaction}, 9(7):1--43, 2025.

\end{thebibliography}

\clearpage

\appendix

\section{Limitations and Future Work}

Our work focuses on suggestion-based interactive completion, which represents an abstraction of the full spectrum of real-world authoring behaviors. Users may engage in a broader range of editing activities—such as reorganizing content, rewriting larger spans, or revisiting earlier decisions—many of which can be decomposed into sequences of localized acceptance and editing actions. By modeling interaction at the level of proactive suggestions and immediate user responses, the proposed HitL-MDP captures a dominant and increasingly common mode of human–AI collaboration in modern writing interfaces.

\paragraph{Limitations.} Nevertheless, several limitations remain. While proactive suggestion-based completion represents the most widely deployed co-writing paradigm today, it is one instantiation of a broader space of human--AI collaborative writing. How the proposed framework and metrics generalize to other collaborative forms---such as outlining assistance or structural critiquing---remains an open question and a natural direction for future investigation.

\paragraph{Future Work.} Looking forward, several directions are worth pursuing. Extending the framework to explicitly represent higher-level revision strategies, longer-horizon planning, or cross-suggestion dependencies would further enrich its descriptive power. Future systems could also be evaluated under more diverse writing conditions, including domain-specific authoring tasks, multilingual settings, and longitudinal user studies that capture how writing preferences evolve over extended interaction. Additionally, studying how HAR and KED generalize to tool-augmented or multi-agent co-writing settings---where writing support is decomposed across complementary roles such as content planning, factual verification, and style refinement---remains a promising direction for interaction-aware evaluation.

\begin{table*}[t]
\centering
\caption{Comparison of acceptance judgment methods for auto-completion suggestions.}
\label{tab:acceptance_comparison}
\small
\begin{tabularx}{\textwidth}{@{} l X X @{}}
\toprule
\textbf{Method} & \textbf{Design Rationale} & \textbf{Empirical Finding} \\
\midrule
\makecell[l]{Logic-Based \\ {\scriptsize(\S\ref{sec:logic-based})}}
& \makecell[l]{Accepts if prediction and reference share \\ the same backbone logical relation (7 categories).}
& \makecell[l]{Over-accepts paraphrastic continuations that \\ are logically valid but add no new information.} \\
\midrule
\makecell[l]{Style-Based \\ {\scriptsize(\S\ref{sec:style-based})}}
& \makecell[l]{Scores four stylistic dimensions (1--10); \\ accepts if surface expression matches.}
& \makecell[l]{Surface match can mask semantic shifts \\ (correlation $\to$ causation), altering the author's claim.} \\
\midrule
\makecell[l]{Semantic-Based \\ {\scriptsize(\S\ref{sec:semantic-based})}}
& \makecell[l]{Measures propositional content overlap (0--10); \\ accepts if the core claim is preserved.}
& \makecell[l]{Insensitive to epistemic stance: hedges \\ (``some studies suggest\ldots'') can be silently dropped.} \\
\midrule
\makecell[l]{Holistic-Based \\ {\scriptsize(\S\ref{sec:holistic})}}
& \makecell[l]{Aggregates entity, logic, style, and semantic \\ signals into one score; accepts if globally similar.}
& \makecell[l]{Single score obscures rejection causes; redundant \\ or uninformative continuations may pass undetected.} \\
\midrule
\makecell[l]{HAR \\ {\scriptsize(\S\ref{sec:har})}}
& \makecell[l]{Hierarchical checklist of interaction, surface, \\ and semantic/intent conditions; all checks must pass.}
& \makecell[l]{Better aligns with human decisions; decomposes \\ acceptance into interpretable stages for ablation.} \\
\bottomrule
\end{tabularx}
\end{table*}

\section{Design of Hierarchical Acceptance Rate}
Human acceptance of auto-completion suggestions is highly selective and asymmetric: some continuations are rejected almost instantly due to obvious violations, while others are evaluated more carefully only after basic interaction constraints are satisfied. As summarized in Table~\ref{tab:acceptance_comparison}, commonly used acceptance proxies---logic-based consistency, stylistic similarity, semantic similarity, and holistic similarity---each exhibit systematic failure modes when applied in isolation: \emph{logic-based judgments} over-accept paraphrastic yet unhelpful continuations; \emph{style-based judgments} overlook semantically consequential shifts; \emph{semantic similarity} fails to account for epistemic stance or key-entity changes; and \emph{holistic similarity} obscures distinct rejection causes by collapsing them into a single score.

To address these mismatches, we conduct a progressive failure analysis, examining representative cases under each criterion before introducing the \emph{Hierarchical Acceptance Rate (HAR)} as an empirically grounded alternative. HAR models acceptance as an ordered checklist of rejection conditions, reflecting how human writers filter suggestions during real-time interaction. Subsequent ablations show that removing or flattening this hierarchy leads to predictable misalignment with human acceptance behavior.

\subsection{Logic-Based Judgment}
\label{sec:logic-based}
\paragraph{Design Rationale.}
The logic-based criterion accepts a continuation if its dominant logical relation---the \emph{backbone logic}---matches that of the reference. Secondary and modifying relations (e.g., descriptive supplements, parenthetical explanations) are ignored; when ambiguous, relations between main clauses and semantically progressive logic (causal, contrastive, sequential, temporal) take priority, with explicit discourse markers serving as strong signals.

We adopt seven backbone categories:
(1) Sequence / Connection,
(2) Causal (including Purpose),
(3) Contrast / Concession,
(4) Elaboration / Explanation,
(5) Temporal,
(6) Frame / Organization,
(7) Other.
A continuation is accepted if its backbone logic category matches that of the reference.

\begin{tcolorbox}[captionbg]
\textbf{Case 1:} Logically valid paraphrase with no informational gain
\end{tcolorbox}

\begin{tcolorbox}[refbox]

\begin{tcolorbox}[referencebg]
\textbf{\textcolor{referencelabelc}{[Reference]}}~~
\emph{In 1642, Blaise Pascal invented the adding machine, the first mechanical calculator, which used interlocking gears to perform addition and subtraction and was mainly applied to practical tasks such as tax calculation.}
\end{tcolorbox}

\begin{tcolorbox}[predictionbg]
\textbf{\textcolor{predictionlabelc}{[Prediction]}}~~ 
\emph{In 1642, Blaise Pascal developed the Pascaline, a gear-driven device capable of eight-digit arithmetic, marking an early step toward the practical use of mechanical computation.}
\end{tcolorbox}

\end{tcolorbox}

\paragraph{Case Study and Discussion.}
Backbone logic provides an effective coarse filter: preserving the dominant logical relation avoids abrupt topic shifts and structural incoherence, making it a reliable signal for rejecting clearly invalid continuations. However, it is inherently insensitive to \emph{informational contribution}. As the Pascal example shows, the prediction follows the same explanatory backbone as the reference yet primarily reformulates existing content rather than advancing the narrative—perceived by human writers as weak assistance not because of logical error, but because of minimal additive value. Logical consistency is thus a necessary but insufficient condition for acceptance: from an interaction perspective, it constrains \emph{how} a continuation can connect, but not \emph{whether it is worth connecting at all}.

\subsection{Style-Based Judgment}
\label{sec:style-based}

\paragraph{Design Rationale.}
The style-based criterion accepts a continuation if it matches the reference in \emph{linguistic style}, independent of factual or semantic content. Style similarity is assessed along four dimensions: lexical style (formality, part-of-speech usage), syntactic structure (sentence length and complexity), linguistic style features (lexical richness, rhetorical devices), and genre/register consistency. Each dimension is scored 1--10 and aggregated via a weighted average; a continuation is accepted if the overall score indicates close surface alignment.

\begin{tcolorbox}[captionbg]
\textbf{Case 2:} Stylistic match masking a semantic shift.
\end{tcolorbox}

\begin{tcolorbox}[refbox]

\begin{tcolorbox}[referencebg]
\textbf{\textcolor{referencelabelc}{[Reference]}}~~
\emph{These results suggest a strong correlation between model capacity and downstream performance.}
\end{tcolorbox}

\begin{tcolorbox}[predictionbg]
\textbf{\textcolor{predictionlabelc}{[Prediction]}}~~
\emph{These results therefore demonstrate that increasing model capacity directly causes improvements in downstream performance.}
\end{tcolorbox}

\end{tcolorbox}

\paragraph{Case Study and Discussion.}
Style-based judgment reliably captures register, sentence structure, and lexical choice, making it useful for assessing fluency and authorial consistency. However, stylistic alignment alone proves insufficient for acceptance. 
As the correlation-versus-causation example shows, the prediction closely mirrors the reference in academic tone and syntax yet subtly strengthens the claim by asserting causality---a shift that is stylistically neutral but semantically consequential, and one that human evaluators typically reject despite apparent fluency. Surface-level similarity can thus obscure changes in meaning or intent, making stylistic consistency a useful compatibility signal rather than a standalone indicator of acceptability.

\subsection{Semantic-Based Judgment}
\label{sec:semantic-based}

\paragraph{Design Rationale.}
The semantic-based criterion accepts a continuation if it preserves the \emph{core propositional content} of the reference (who/what/when/where/why/how), independent of surface form or discourse structure. This design reflects the intuition that human writers routinely accept paraphrases and syntactic rewrites as long as the underlying meaning is intact---making surface variation a poor grounds for rejection. Semantic similarity is scored 0--10, where paraphrasing or bidirectional entailment yields high scores and partial overlap, scope shifts, or modality changes incur deductions; the fine-grained scale is intended to handle borderline cases such as elaborations or partial paraphrases with appropriate nuance. A continuation is accepted if it preserves the factual scope, polarity, and epistemic strength of the reference claim.

\begin{tcolorbox}[captionbg]
\textbf{Case 3:} Hedge removal altering epistemic stance
\end{tcolorbox}

\begin{tcolorbox}[refbox]

\begin{tcolorbox}[referencebg]
\textbf{\textcolor{referencelabelc}{[Reference]}}~~
\emph{Some studies suggest that vibe coding reduces cognitive load.}
\end{tcolorbox}

\begin{tcolorbox}[predictionbg]
\textbf{\textcolor{predictionlabelc}{[Prediction]}}~~
\emph{Vibe coding reduces cognitive load.}
\end{tcolorbox}

\end{tcolorbox}

\paragraph{Case Study and Discussion.}
Semantic similarity effectively detects paraphrases and preserves core factual content, making it reliable when entities, outcomes, and topical focus align. However, it is insensitive to \emph{epistemic stance}. In the hedging example, removing ``some studies suggest’’ presents a qualified claim as an unqualified fact: the core proposition is unchanged, but the author’s intended level of certainty is altered---a difference salient to human evaluators but invisible to similarity metrics. High semantic overlap therefore does not guarantee intent preservation; by collapsing hedged and definitive assertions into a single continuum, semantic criteria may systematically over-accept continuations that preserve content while shifting epistemic commitment.

\subsection{Holistic Judgment}
\label{sec:holistic}

\paragraph{Design Rationale.}
The holistic criterion accepts a continuation if it is deemed globally consistent with the reference across four aspects: entity similarity, logical similarity, stylistic similarity, and semantic similarity. The appeal of this design lies in its resemblance to how a reader might form an overall impression: rather than auditing each dimension independently, a single integrative score captures the joint contribution of all factors. Unlike the preceding single-dimension judgments, it implicitly synthesizes multiple signals into one acceptance decision---reflecting common practice in prior work that treats similarity as a weighted or intuitive aggregation rather than a structured process. However, collapsing heterogeneous signals into one scalar inevitably discards the relative weighting and ordering information that distinguishes, for instance, a logical inconsistency from a mere stylistic mismatch.

\begin{tcolorbox}[captionbg]
\textbf{Case 4:} Globally similar prediction that redundantly repeats the immediate context
\end{tcolorbox}

\begin{tcolorbox}[refbox]

\begin{tcolorbox}[contextbg]
\textbf{\textcolor{contextlabelc}{[Context]}}~~
\emph{Although it may}
\end{tcolorbox}

\begin{tcolorbox}[referencebg]
\textbf{\textcolor{referencelabelc}{[Reference]}}~~
\emph{appear quite cumbersome by contemporary standards, it constituted the most expeditious computational method of its era.}
\end{tcolorbox}

\begin{tcolorbox}[predictionbg]
\textbf{\textcolor{predictionlabelc}{[Prediction]}}~~
\emph{Although it may appear quite primitive in retrospect, it constituted the most expeditious computational method of its era.}
\end{tcolorbox}

\end{tcolorbox}

\vspace{6pt}

\begin{tcolorbox}[captionbg]
\textbf{Case 5:} Factually correct but minimally informative fragment in list completion
\end{tcolorbox}

\begin{tcolorbox}[refbox]

\begin{tcolorbox}[contextbg]
\textbf{\textcolor{contextlabelc}{[Context]}}~~
\emph{8. Transfer clips, two sponge pads, a dropper, filter paper, and a sheet of}
\end{tcolorbox}

\begin{tcolorbox}[referencebg]
\textbf{\textcolor{referencelabelc}{[Reference]}}~~
\emph{PVDF membrane, transfer tank, electrophoretic transfer apparatus, shaker}
\end{tcolorbox}

\begin{tcolorbox}[predictionbg]
\textbf{\textcolor{predictionlabelc}{[Prediction]}}~~
\emph{PVDF membrane.}
\end{tcolorbox}

\end{tcolorbox}


\paragraph{Case Study and Discussion.}
Holistic judgment provides a convenient summary signal when multiple dimensions align, often agreeing with human intuition. However, aggregating heterogeneous factors into a single score obscures interaction-specific constraints central to auto-completion. The prefix-repetition case illustrates this: the prediction is globally similar to the reference yet repeats the immediate context, disrupting writing flow in a way holistic scoring cannot detect. Similarly, listing ``PVDF membrane'' without advancing the sequence is factually correct and globally consistent, yet offers little assistive value. By collapsing distinct failure modes into one decision, holistic judgment lacks the resolution to identify \emph{why} a continuation is unhelpful---leading to inconsistent acceptance of redundant or minimally informative completions without interpretable explanation.

\subsection{Hierarchical Acceptance Rate}
\label{sec:har}
\paragraph{Design Rationale.}
The preceding analyses collectively show that human acceptance cannot be reduced to a single similarity signal or an unstructured aggregation of multiple criteria. Across cases, acceptance emerges as a \emph{hierarchical decision process}: certain violations trigger immediate rejection, while others become relevant only after more fundamental constraints are satisfied.

HAR operationalizes this as a conservative, rule-based protocol in which a continuation is accepted only if it passes a sequence of ordered block conditions---structured from low-level, immediately perceptible violations to high-level, intent-sensitive judgments. Over-accepting weak or misleading completions is often more disruptive than rejecting a borderline suggestion, motivating the strictness of this design.

The checklist is organized into three layers:

\begin{itemize}
  \item \textbf{Layer 1 --- Interaction Constraints.} Fundamental validity checks that users typically reject without considering content: start repetition, language mismatch, semantic incoherence, and structural closure (paired punctuation, Markdown/\LaTeX{}/code blocks).

  \item \textbf{Layer 2 --- Surface \& Format Compatibility.} Checks for continuations that are locally valid but disrupt document progression: format mismatch, inconsistency with preceding text, depth mismatch, style/register mismatch, sentence type inconsistency, and perspective shift.

  \item \textbf{Layer 3 --- Semantic \& Intent.} Higher-order conditions evaluated only after Layers 1--2 pass: early semantic overlap, subset acceptance for lists/tables, topic divergence, missing or substituted key entities, and intent mismatch. A final comprehensive judgment resolves residual edge cases.
\end{itemize}

This hierarchical structure decomposes acceptance into interpretable stages, supports stable evaluation across models, and enables ablation analysis over rule subsets.

\newcommand{\rj}{\textcolor{red!65!black}{\textit{$\to$ Reject}}}
\newcommand{\ac}{\textcolor{green!50!black}{\textit{$\to$ Accept}}}
\newcommand{\acrej}{\textcolor{black!50}{\textit{$\to$ Accept / Reject}}}

\begin{algorithm}[t]
\small
\renewcommand{\algorithmicrequire}{\textbf{Input:}}
\renewcommand{\algorithmicensure}{\textbf{Output:}}
\caption{Hierarchical Acceptance Rate (HAR)}
\label{alg:har}
\begin{algorithmic}[1]
\REQUIRE \texttt{USER\_INPUT} ($U$),\; \texttt{COMPLETION} ($C$),\; \texttt{REFERENCE} ($R$)
\ENSURE \textsc{Accept} or \textsc{Reject}
\STATE \textit{$\triangleright$ \textbf{Layer 1: Interaction Constraints}}
\STATE \textbf{if} $C$ starts with repetition of $U$ \textbf{then return} \textsc{Reject}
\STATE \textbf{if} $\mathrm{lang}(C) \neq \mathrm{lang}(R)$ \textbf{then return} \textsc{Reject}
\STATE \textbf{if} $U \oplus C$ is semantically incoherent \textbf{then return} \textsc{Reject}
\STATE \textbf{if} $\mathrm{early\_overlap}(C, R) > 50\%$ \textbf{then return} \textsc{Accept}
\STATE \textbf{if} paired punctuation opened in $U$ is unclosed in $C$ \textbf{then return} \textsc{Reject}
\STATE \textbf{if} Markdown/\LaTeX{}/code block opened in $U$ is unclosed in $C$ \textbf{then return} \textsc{Reject}
\STATE \textit{$\triangleright$ \textbf{Layer 2: Surface \& Format Compatibility}}
\STATE \textbf{if} $\mathrm{format}(C) \neq \mathrm{format}(R)$ \textbf{then return} \textsc{Reject}
\STATE \textbf{if} format of $C$ diverges from preceding text in $U$ \textbf{then return} \textsc{Reject}
\STATE \textbf{if} $|\mathrm{depth}(C) - \mathrm{depth}(R)| > 30\%$ \textbf{then return} \textsc{Reject}
\STATE \textbf{if} style/register of $C$ diverges from $R$ \textbf{then return} \textsc{Reject}
\STATE \textbf{if} sentence type of $C$ differs from $R$ \textbf{then return} \textsc{Reject}
\STATE \textbf{if} narrative perspective of $C$ shifts from $U$ \textbf{then return} \textsc{Reject}
\STATE \textit{$\triangleright$ \textbf{Layer 3: Semantic \& Intent}}
\STATE \textbf{if} $C$ is a true subset of $R$ (list/table) \textbf{then return} \textsc{Accept}
\STATE \textbf{if} $\mathrm{topic}(C) \neq \mathrm{topic}(R)$ \textbf{then return} \textsc{Reject}
\STATE \textbf{if} key entities in $C$ missing or altered vs.\ $R$ \textbf{then return} \textsc{Reject}
\STATE \textbf{if} $\mathrm{intent}(C) \neq \mathrm{intent}(R)$ \textbf{then return} \textsc{Reject}
\RETURN $\mathrm{comprehensive\_judgment}(C, R)$
\end{algorithmic}
\end{algorithm}

\paragraph{Motivation for Layer-wise Analysis.}
While HAR is designed as a complete hierarchical protocol,
its structure naturally admits meaningful subsets of conditions.
In particular, early-stage checks capture fundamental interaction validity,
whereas later-stage rules model increasingly subtle aspects of writing quality
and intent preservation.
To better understand the contribution of different stages,
we conduct a layer-wise analysis of HAR
and examine representative cases where partial checklists fail.

The following cases illustrate complementary limitations:
the first highlights failure modes that persist
when only early-stage conditions are applied (Layer1),
while the second demonstrates errors that remain undetected
even after incorporating additional stylistic and semantic checks (Layer1+2).
Rather than serving as isolated counterexamples,
these cases reflect recurrent patterns observed across our evaluation data,
and clarify why a full hierarchical protocol is necessary.

\begin{tcolorbox}[captionbg]
\textbf{Case 6:} Depth mismatch --- abstract completion following a concrete experimental setup (Layer~1 failure)
\end{tcolorbox}

\begin{tcolorbox}[refbox]

\begin{tcolorbox}[contextbg]
\textbf{\textcolor{contextlabelc}{[Context]}}~~
\emph{To evaluate the catalytic activity of the synthesized material, we conducted a series of controlled experiments.}
\end{tcolorbox}

\begin{tcolorbox}[referencebg]
\textbf{\textcolor{referencelabelc}{[Reference]}}~~
\emph{Specifically, the reaction yield was measured at 300 K under 1 atm using gas chromatography.}
\end{tcolorbox}

\begin{tcolorbox}[predictionbg]
\textbf{\textcolor{predictionlabelc}{[Prediction]}}~~
\emph{The catalytic performance was evaluated using standard experimental procedures.}
\end{tcolorbox}

\end{tcolorbox}

\paragraph{Case Study: Depth Mismatch Beyond Surface Validity.}
This completion satisfies Layer~1 constraints and is grammatically and semantically plausible, so a checklist limited to early-stage conditions would not reject it. However, scientific experimental writing typically progresses from high-level setup to concrete parameters; the completion retreats to a more abstract level, disrupting the expected depth progression. Layer~2's explicit depth consistency check detects this form of interference, which eludes surface-level validity checks alone.

\begin{tcolorbox}[captionbg]
\textbf{Case 7:} Entity-level drift --- XRD substituted by SEM despite different scientific functions (Layer~1+2 failure)
\end{tcolorbox}

\begin{tcolorbox}[refbox]

\begin{tcolorbox}[contextbg]
\textbf{\textcolor{contextlabelc}{[Context]}}~~
\emph{The crystal structure of the synthesized samples was analyzed.}
\end{tcolorbox}

\begin{tcolorbox}[referencebg]
\textbf{\textcolor{referencelabelc}{[Reference]}}~~
\emph{X-ray diffraction (XRD) patterns confirmed the formation of the target phase.}
\end{tcolorbox}

\begin{tcolorbox}[predictionbg]
\textbf{\textcolor{predictionlabelc}{[Prediction]}}~~
\emph{Scanning electron microscopy (SEM) images confirmed the formation of the target phase.}
\end{tcolorbox}

\end{tcolorbox}

\paragraph{Case Study: Entity-Level Drift Beyond Semantic Consistency.}
This completion is fluent, thematically aligned, and exhibits high semantic similarity---both sentences describe a technique confirming a target phase---so even a Layer~1+2 checklist may not flag it. Yet XRD and SEM serve fundamentally different functions: XRD determines crystal structure while SEM characterizes surface morphology. Substituting one for the other is \emph{entity-level drift} that is subtle yet scientifically incorrect, and undetectable without explicit key-entity and intent checks. Layer~3 provides precisely this resolution, distinguishing superficially plausible continuations from those that violate domain-specific expectations.

\section{Design of Knowledge-aware Editing Distance.}
In this appendix, we present qualitative case studies illustrating the limitations of conventional distance and similarity metrics in estimating human editing effort. Each case demonstrates a mismatch between metric scores and actual post-editing cost, as observed in human revision behavior. Table~\ref{tab:ked_comparison} summarizes these comparisons; the following cases contrast these failures with KED, which accounts for the cognitive cost of correcting semantic, factual, and structural inconsistencies.

\begin{table*}[t]
\centering
\caption{Comparison of editing cost metrics for human revision effort estimation.}
\label{tab:ked_comparison}
\small
\begin{tabularx}{\textwidth}{@{} l X X @{}}
\toprule
\textbf{Method} & \textbf{Design Rationale} & \textbf{Empirical Finding} \\
\midrule
\makecell[l]{Levenshtein Distance \\ {\scriptsize(\hyperref[sec:ked-lev]{Case 8})}}
& \makecell[l]{Character-level edit distance; assigns uniform cost to \\ all substitutions regardless of semantic significance.}
& \makecell[l]{Underestimates factual repair cost; treats entity changes \\ (273\,K\,$\to$\,298\,K) as trivially low-cost edits.} \\
\addlinespace
\makecell[l]{Compression Distance \\ {\scriptsize(\hyperref[sec:ked-comp]{Case 9})}}
& \makecell[l]{LZ77-based string reuse; distance captures syntactic \\ redundancy but lacks semantic entity awareness.}
& \makecell[l]{Insensitive to method-level substitutions; assigns near-zero \\ cost to swapping qRT-PCR with Western blotting.} \\
\addlinespace
\makecell[l]{Revision Distance \\ {\scriptsize(\hyperref[sec:ked-rev]{Case 10})}}
& \makecell[l]{Counts edit operations (ADD, DELETE, REPLACE); \\ each step is treated as equally costly.}
& \makecell[l]{Misses reasoning-level cost; a single ADD introducing \\ a conditional constraint appears as a low-cost edit.} \\
\addlinespace
\makecell[l]{ROUGE / BLEU \\ {\scriptsize(\hyperref[sec:ked-rouge]{Case 11})}}
& \makecell[l]{N-gram overlap metrics; reward high lexical similarity \\ between prediction and reference.}
& \makecell[l]{Fail to penalize overstated claims; treat ``competitive'' \\ $\leftrightarrow$``state-of-the-art'' as near-identical.} \\
\addlinespace
\makecell[l]{BERTScore \\ {\scriptsize(\hyperref[sec:ked-bert]{Case 12})}}
& \makecell[l]{Contextual embedding similarity; measures semantic \\ proximity between tokens.}
& \makecell[l]{Insensitive to causal direction errors; assigns high \\ similarity to sentences with reversed causal logic.} \\
\addlinespace
\makecell[l]{\textbf{KED (Ours)}}
& \makecell[l]{Decomposes edits by entity complexity, relation, \\ and semantic role; costs reflect cognitive demands.}
& \makecell[l]{Better aligns with human editing effort; captures \\ factual, methodological, and causal correction costs.} \\
\bottomrule
\end{tabularx}
\end{table*}

\paragraph{Levenshtein Distance Underestimates Semantic Repair Cost.}
\label{sec:ked-lev}
Character-level edit distance assigns uniform cost to all substitutions, failing to distinguish factual entity corrections from trivial character changes.

\begin{tcolorbox}[captionbg]
\textbf{Case 8:} Factual entity change (temperature) underestimated by character-level distance
\end{tcolorbox}
\begin{tcolorbox}[refbox]

\begin{tcolorbox}[referencebg]
\textbf{\textcolor{referencelabelc}{[Reference]}}~~
\emph{The experiment was conducted at \underline{298 K} under standard atmospheric pressure.}
\end{tcolorbox}

\begin{tcolorbox}[predictionbg]
\textbf{\textcolor{predictionlabelc}{[Prediction]}}~~
\emph{The experiment was conducted at \underline{273 K} under standard atmospheric pressure.}
\end{tcolorbox}

\end{tcolorbox}

Levenshtein treats the substitution ``273\,K\,$\to$\,298\,K'' as a low-cost edit involving only a few character changes. In scientific writing, however, experimental temperature is a factually critical entity: correcting it requires identifying a measurement error and cross-checking laboratory records or domain knowledge---a cognitively demanding operation, not a typographical fix. KED classifies this as a complex entity modification and assigns a correspondingly higher cost, better reflecting actual human editing effort.

\paragraph{Compression Distance Is Insensitive to Method-Level Semantic Errors.}
\label{sec:ked-comp} Compression-based distances derived from LZ77 primarily capture string reuse and lack semantic awareness of key entity substitutions.

\begin{tcolorbox}[captionbg]
\textbf{Case 9:} Method-level substitution undetected by compression-based distance
\end{tcolorbox}
\begin{tcolorbox}[refbox]

\begin{tcolorbox}[referencebg]
\textbf{\textcolor{referencelabelc}{[Reference]}}~~
\emph{We employed \underline{qRT-PCR} to quantify RNA expression levels.}
\end{tcolorbox}

\begin{tcolorbox}[predictionbg]
\textbf{\textcolor{predictionlabelc}{[Prediction]}}~~
\emph{We employed \underline{Western blotting} to quantify RNA expression levels.}
\end{tcolorbox}

\end{tcolorbox}

The two sentences share nearly identical syntactic structure and lexical overlap, yielding a minimal Compression Distance. Yet ``Western blotting'' and ``qRT-PCR'' are mutually exclusive techniques---the former targets proteins, the latter RNA---making this a domain-critical substitution that demands methodological knowledge to detect and correct. KED treats the method name as a complex entity and additionally accounts for the predicate-level semantic correction, providing a more accurate cost estimate.

\paragraph{Revision Distance Fails to Capture the Cost of Reasoning and Constraint Introduction.}
\label{sec:ked-rev} Edit-step–based distances overlook the cognitive planning involved in a single revision.

\begin{tcolorbox}[captionbg]
\textbf{Case 10:} Reasoning constraint introduction counted as a single low-cost ADD step
\end{tcolorbox}
\begin{tcolorbox}[refbox]

\begin{tcolorbox}[referencebg]
\textbf{\textcolor{referencelabelc}{[Reference]}}~~
\emph{These results suggest that the model improves performance \underline{only under low-noise conditions.}}
\end{tcolorbox}

\begin{tcolorbox}[predictionbg]
\textbf{\textcolor{predictionlabelc}{[Prediction]}}~~
\emph{These results suggest that the model improves performance.}
\end{tcolorbox}

\end{tcolorbox}

Revision Distance counts this as a single ADD operation and assigns low cost. In reality, the added phrase introduces a critical conditional constraint that substantially narrows the claim's scope---requiring the editor to reassess experimental conditions, validate the conclusion boundary, and check consistency with earlier settings. This is a reasoning-level correction, not a surface edit. KED accounts for both the complex constraint entity and the structural modification to the relational phrasing, capturing the true cognitive burden of the revision.

\paragraph{ROUGE and BLEU Fail to Penalize Overstated Scientific Claims.}
\label{sec:ked-rouge} N-gram–based similarity metrics reward lexical overlap while ignoring changes in the strength of scientific claims.

\begin{tcolorbox}[captionbg]
\textbf{Case 11:} Claim strength change ignored by n-gram overlap
\end{tcolorbox}
\begin{tcolorbox}[refbox]

\begin{tcolorbox}[referencebg]
\textbf{\textcolor{referencelabelc}{[Reference]}}~~
\emph{The algorithm achieves \underline{competitive} performance on the benchmark.}
\end{tcolorbox}

\begin{tcolorbox}[predictionbg]
\textbf{\textcolor{predictionlabelc}{[Prediction]}}~~
\emph{The algorithm achieves \underline{state-of-the-art} performance on the benchmark.}
\end{tcolorbox}

\end{tcolorbox}

High lexical overlap leads ROUGE-L and BLEU to assign near-identical scores, yet ``state-of-the-art'' and ``competitive'' represent substantially different levels of scientific assertion---a distinction that can affect the abstract, conclusions, and submission positioning. KED treats this as a claim-strength modification and assigns a cost that reflects the associated cognitive and rhetorical effort.

\paragraph{BERTScore Fails to Detect Errors in Causal Direction.}
\label{sec:ked-bert} Embedding-based similarity metrics are insensitive to changes in the direction of logical relations.

\begin{tcolorbox}[captionbg]
\textbf{Case 12:} Causal Reversal Missed by Embedding Similarity
\end{tcolorbox}
\begin{tcolorbox}[refbox]

\begin{tcolorbox}[referencebg]
\textbf{\textcolor{referencelabelc}{[Reference]}}~~
\emph{This improvement \underline{enables} the model to scale to larger capacities.}
\end{tcolorbox}

\begin{tcolorbox}[predictionbg]
\textbf{\textcolor{predictionlabelc}{[Prediction]}}~~
\emph{This improvement \underline{is caused by} the increase in model capacity.}
\end{tcolorbox}

\end{tcolorbox}

Shared entities yield a high BERTScore despite a reversed causal direction---``enables'' (A causes B) versus ``is caused by'' (B causes A)---which alters the backbone logic of the argument. Correcting such errors requires reexamining the underlying reasoning structure, a cognitively demanding task that surface similarity cannot capture. KED treats causal direction reversal as a high-cost relational restructuring, more faithfully reflecting human editing effort.

\section{Dataset and Full Scores}

\paragraph{Dataset Composition.}
The evaluation dataset consists of 60 human-authored articles drawn from 16 writing domains, covering both Scientific and Creative writing scenarios. Scientific domains emphasize factual correctness, domain-specific entities, and procedural precision, while Creative domains allow greater stylistic freedom and semantic flexibility.

\paragraph{Article Coverage.}
Table~\ref{tab:dataset_1} and Table~\ref{tab:dataset_2} list all domains included in the dataset, along with the corresponding article titles and publication dates. The selected articles span a wide range of topics, including technical reports, experimental protocols, software analysis, and narrative-style expository writing, and collectively cover a broad temporal range from classical literary works to contemporary scientific publications.

\paragraph{Query Construction.}
For each article, multiple continuation points are sampled from early, middle, and late sections to capture different stages of document development. At each point, the preceding context defines the article state, and the original text following that point serves as the ground-truth continuation. This ground truth is shared across all interaction paradigms to ensure controlled comparison. In total, this process yields 1,688 evaluation queries.

\paragraph{Model Full Scores.}
Table~\ref{tab:gpt51_constraint_domains} reports full per-domain HAR and KED scores for all evaluated models under both L1 and L2 interaction settings, covering 16 fine-grained writing domains across Scientific and Creative categories. Domain-specific results reveal how model behavior varies under different interaction constraints and writing scenarios.

\section{System Design of Adaptive Co-Writing}
\label{sec:system_design}

The L3 interaction paradigm is instantiated as an adaptive proactive co-writing assistant that operationalizes the \emph{Human-in-the-Loop MDP} formulation, translating empirical insights from L1 and L2 evaluations into concrete design choices for intent modeling and intervention timing.

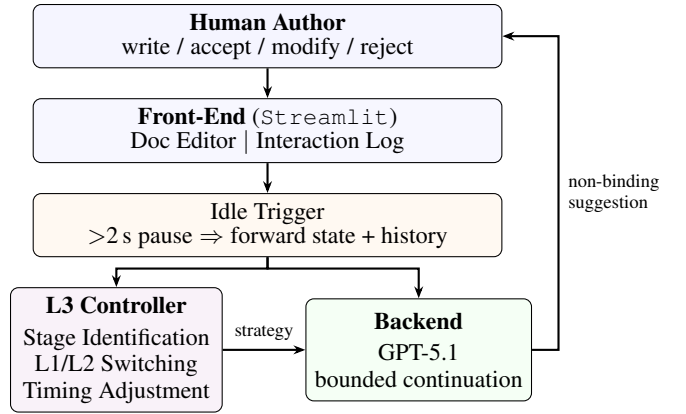
\begin{figure}[t]
\centering
\begin{tikzpicture}[
  arr/.style={-{Stealth[length=4.5pt]}, thick},
  mbox/.style={draw, rounded corners=3pt, fill=#1,
               minimum width=6.2cm, minimum height=0.62cm,
               align=center, font=\footnotesize},
  hbox/.style={draw, rounded corners=3pt, fill=#1,
               minimum width=2.8cm, minimum height=1.35cm,
               align=center, font=\footnotesize},
]

\node[mbox=blue!4] (author) at (0, 0)
  {\textbf{Human Author}\\[-1pt]write\;/\;accept\;/\;modify\;/\;reject};

\node[mbox=blue!3] (frontend) at (0,-1.25)
  {\textbf{Front-End} (\texttt{Streamlit})\\[-1pt]
   Doc Editor\;\textbar\;Interaction Log};

\node[mbox=orange!5] (trigger) at (0,-2.50)
  {Idle Trigger\\[-1pt]$>$2\,s\;pause\;$\Rightarrow$\;forward state + history};

\node[hbox=violet!5] (ctrl) at (-2.0,-4.15)
  {\textbf{L3 Controller}\\[2pt]
   Stage Identification\\L1/L2 Switching\\Timing Adjustment};

\node[hbox=green!4] (backend) at (2.0,-4.15)
  {\textbf{Backend}\\[2pt]
   GPT-5.1\\bounded continuation};

\draw[arr] (author)    -- (frontend);
\draw[arr] (frontend)  -- (trigger);
\draw[arr] (trigger.south) -- ++(0,-0.15) -| (ctrl.north);
\draw[arr] (trigger.south) -- ++(0,-0.15) -| (backend.north);
\draw[arr] (ctrl.east) -- (backend.west)
  node[midway, above, font=\scriptsize]{strategy};
\draw[arr] (backend.east) -- ++(0.35,0)
           |- (author.east)
  node[pos=0.25, right, font=\scriptsize, align=left]{non-binding\\suggestion};

\end{tikzpicture}
\caption{Architecture of the adaptive proactive co-writing system.}
\label{fig:system_arch}
\end{figure}

\paragraph{System Overview.}
The system adopts a client--server architecture comprising a lightweight front-end writing interface and a GPT-5.1-powered backend assistant (Figure~\ref{fig:system_arch}). It supports suggestion-based interactive completion, where the assistant proactively proposes localized continuations while all document updates remain under full human control. Unlike fixed proactive systems, the L3 design treats intervention behavior as a policy-level decision: both the intent modeling strategy and the intervention timing are dynamically adapted based on the inferred writing stage.

\paragraph{Front-End Interface and Event Triggering.}
The front-end is implemented in \texttt{Streamlit}, providing a document-centric editor with minimal interaction overhead. Beyond document content, the interface maintains a structured interaction log recording prior suggestions and corresponding user responses (acceptance, modification, or rejection).

Suggestion delivery is governed by an idle-based triggering mechanism: when no textual change is detected for more than 2 seconds, the current document state and interaction history are forwarded to the backend. Short typing pauses serve as implicit signals for assistance, enabling proactive support without requiring explicit user prompts.

\paragraph{Backend Writing Assistant.}
The backend conditions on (i) the current document state $s_t$, (ii) the accumulated interaction history, and (iii) the interaction strategy selected by the adaptive controller. GPT-5.1 generates a bounded, localized continuation aligned with surrounding context in scope, style, and length. Consistent with the action space $\mathcal{A}$, all suggestions are non-binding: surfaced as candidate text, with any document update realized solely through subsequent human action.

\paragraph{Adaptive L3 Policy.}
The defining feature of the L3 system is an adaptive intervention policy that jointly determines \emph{how} suggestions are generated and \emph{when} they are delivered:

\begin{itemize}
  \item \textbf{Writing Stage Identification.} An auxiliary lightweight agent infers the current writing stage from document-level signals---text length, structural cues, and recent interaction patterns. The inferred stage drives system control internally and is not exposed to the user.

  \item \textbf{Strategy Switching Between L1 and L2.} Experimental results show that stateless suggestions (L1) achieve higher acceptance during early writing stages when author intent is still forming, while stateful suggestions (L2), conditioning on accumulated interaction history, outperform in later stages. The L3 system therefore applies L1-style stateless mode in early stages and switches to L2-style stateful mode as writing progresses.

  \item \textbf{Adaptive Intervention Timing.} The intervention delay is adjusted dynamically: longer idle thresholds are applied in early stages to reduce interruption when acceptance rates are lower, and shorter thresholds in later stages where suggestions better align with stabilized author intent.
\end{itemize}

\section{Details of User Study}

\paragraph{Participant Recruitment.}
Thirty participants were recruited from universities and research institutions, comprising 25 males and 5 females. Nineteen held or were actively pursuing a doctoral degree, reflecting a participant pool with substantial academic writing experience. All participants were native or proficient users of the target writing language and had prior experience with AI-assisted writing tools.

\paragraph{Questionnaire Design.}
After each writing task, participants completed a concise post-task questionnaire designed to capture their experience across different interactive co-writing paradigms. As shown in Table~\ref{tab:user_study_questionnaire}, the questionnaire covers four complementary dimensions: Perceived Workload, Usefulness, Customization, and Adaptability. Perceived Workload follows the raw NASA-TLX dimensions to assess cognitive cost independently of system quality, while the remaining dimensions capture the contextual usefulness of suggestions, the emergence of personalization over interaction history, and users’ perceived appropriateness and controllability of system interventions. All items are rated on a 7-point Likert scale, ranging from \emph{Very Bad} (1) to \emph{Very Good} (7).

\paragraph{Score Distribution.}
The full distribution of user ratings across all interaction paradigms and questionnaire items is reported in Table~\ref{tab:user_study_scores_full}, providing a quantitative basis for analyzing how different levels of proactivity and adaptivity shape user experience.

\section{Details of Training}

\label{sec:appendix_rl}

This subsection supplements the main-text section on learning with interaction-aware rewards by detailing the concrete training setup and data construction.

\paragraph{Training Data Construction.}
Training Data Construction. To support stable behavior-grounded learning, we construct a training corpus directly from the user study interaction logs. Specifically, we leverage the writing session logs collected during the user study, where each session records the full sequence of AI suggestions alongside the user's final decision. For each logged interaction, the preceding document context serves as the input, and the user's accepted suggestion or final edited text serves as the gold reference continuation. This data construction strategy ensures that the reward signal is grounded in authentic human revision behavior rather than in proxy annotations. The resulting dataset is used exclusively for reinforcement learning and is disjoint from all evaluation sets.

\paragraph{Reward Design.} The reward signal is derived from HAR: given a generated continuation and the gold reference from the training corpus, HAR evaluates whether the continuation satisfies all hierarchical acceptance conditions. A continuation that passes all HAR validators receives a positive reward, while one that fails any condition receives no reward. This directly aligns the training objective with the interaction-aware acceptance criterion used at evaluation time.

\paragraph{Training Setup.}
We adopt \emph{Group Relative Policy Optimization (GRPO)} as the reinforcement learning algorithm. GRPO optimizes relative preferences among multiple rollouts and has been shown to be effective in stabilizing learning under noisy or implicit reward signals. We use a batch size of 256 and sample $n_{\text{rollout}} = 8$ candidate continuations per query. Training is conducted under two interaction paradigms: L1, where suggestions are generated without interaction history, and L2, where the model conditions on accumulated interaction trajectories.

\section{Prompt Templates}

All prompt templates used in the experiments are listed at the end of paper, organized by function.
Each template is designed to elicit a structured, reproducible judgment from the underlying language model, with explicit scoring rubrics and output format constraints to minimize ambiguity and variance across runs.
Five prompts each operationalize a distinct acceptance judgment criterion: \textbf{HAR} (rule-based hierarchical checklist), \textbf{logic-based} (backbone logical relationship), \textbf{style-based} (style similarity), \textbf{semantic-based} (semantic similarity), and \textbf{holistic-based} (joint four-dimension scoring).
The \textbf{Edit\_Distance} prompt implements the editing cost assessment for the KED metric.
Finally, \textbf{Completion\_L1} and \textbf{Completion\_L2} are the system prompts for the stateless and stateful writing continuation assistants, respectively; Completion\_L1 operates without document-level context, while Completion\_l2 conditions on the full preceding passage to support coherent long-form generation.

\clearpage

\begin{table*}[t]
\centering
\caption{Articles included in the \textbf{Scientific domain category}, covering technical reports, experimental protocols, and other fact-constrained writing. Each entry lists the article title and publication date.}
\label{tab:dataset_1}
\small
\setlength{\tabcolsep}{4pt}
\begin{tabularx}{\textwidth}{
    c
    >{\raggedright\arraybackslash}X
    c
}
\toprule
\textbf{ID} & \textbf{Title} & \textbf{Publication Date} \\
\midrule

\rowcolor{gray!15}
\multicolumn{3}{l}{\textit{(D1) Technique Report}} \\
1  & Western Blot: Principles, Experimental Workflow, and Common Issues & 2020-05-16 \\
2  & FLUX.1 Source Code Analysis & 2024-09-28 \\
3  & Three Ways to Deploy the DeepSeek-R1 Large Model Locally---One for Every Need & 2025-02-03 \\
4  & Molecular Experiments: Micropipetting and Micro-Operation Techniques & 2025-07-25 \\
5  & Nuclear Magnetic Resonance (NMR) Spectroscopy: Principles, Applications, Analysis, and Common Issues & 2019-08-15 \\

\midrule
\rowcolor{gray!15}
\multicolumn{3}{l}{\textit{(D2) Paper Reading}} \\
6  & A Less-Discussed Autoregressive Generative Model: An In-Depth Look at the PixelCNN Family & 2023-05-27 \\
7  & Toward Proactive, Self-Evolving, and Privacy-Preserving Intelligent Personal Assistants: A Deep Dive into the Galaxy Framework & 2025-08-28 \\
8  & Cellular ``Civil War'': Mitochondria Enforce a Scorched-Earth Strategy to Starve Invading Parasites & 2025-08-24 \\

\midrule
\rowcolor{gray!15}
\multicolumn{3}{l}{\textit{(D3) Survey Report}} \\
9  & Agent Optimization Paradigms & 2025-08-13 \\
10 & Economic Performance Analysis of Shanghai's Nonferrous Metals Industry in the First Half of 2025 & 2025-08-26 \\
11 & Global Automotive Market Analysis in the First Half of 2025 and Full-Year Outlook & 2025-08-29 \\
12 & Interpreting Cancer Statistics 2025: A Comparative Study of Epidemiological Characteristics and Long-Term Trends in China and the U.S. & 2025-03-18 \\
13 & China's Phased Carbon Emission Peaks: What Do They Mean? & 2025-07-18 \\
14 & New Advances in Norovirus Research in 2025: From Pathogenesis to Prevention and Control & 2025-03-22 \\
15 & In-Depth Analysis of Provincial GDP Data in the First Half of 2025 & 2025-08-26 \\
16 & Gas Market Report, Q3-2025 & 2025-07-22 \\
17 & Progress in Video Generation Model Technology & 2025-06-15 \\

\midrule
\rowcolor{gray!15}
\multicolumn{3}{l}{\textit{(D4) Popular Science Article}} \\
18 & The Maillard Reaction: A ``Hot Romance'' Between Amino Acids and Sugars & 2025-08-28 \\
19 & What Protects Earth's Atmosphere from Being Stripped Away by the Solar Wind? & 2024-03-19 \\
20 & The Evolution of Brain Complexity: A Story of Space, Time, and Entropy & 2024-03-31 \\
21 & An Introduction to Earth's Atmosphere and Its Vertical Structure & 2020-01-15 \\

\midrule
\rowcolor{gray!15}
\multicolumn{3}{l}{\textit{(D5) Principle Analysis}} \\
22 & Does Disabling Secure Boot Alleviate the CA Certificate Bottleneck? Reexamining the Secure Boot Constraint & 2025-07-15 \\
23 & Constrained Optimization and Steepest Descent: From SGD to Optimization on the Hypersphere & 2025-08-01 \\
24 & A Deep Analysis of the Major Factors Influencing Animal Immunity & 2025-08-27 \\
25 & From Prompt to Context: Why Think Tools Are an Inevitable Formalization & 2025-08-20 \\
26 & Chang'e-6 Unveils the Mystery of ``Weak Magnetic Fields but Strong Magnetism'' on the Far Side of the Moon & 2025-07-09 \\
27 & Structured Light: From Smartphone Face Unlocking to Digital Twins & 2025-08-28 \\
28 & From RAG to Context Engineering: Redefining the Cognitive Boundaries of AI Systems & 2025-09-01 \\
29 & Often-Overlooked Issues in Assessing Aromaticity Using NICS and Induced Ring Currents & 2025-06-11 \\
30 & From Cursor to Claude Code: Eliminating AI Coding Agent Configuration Hassles and a Detailed Introduction to ASL, the Next-Generation Universal Agent Specification Language & 2025-08-27 \\

\bottomrule
\end{tabularx}
\end{table*}

\clearpage

\begin{table*}[t]
\centering
\caption{Articles included in the \textbf{Creative domain category}, representing writing with greater stylistic flexibility and interpretive latitude. Each entry lists the article title and publication date.}
\label{tab:dataset_2}
\small
\setlength{\tabcolsep}{4pt}
\begin{tabularx}{\textwidth}{
    c
    >{\raggedright\arraybackslash}X
    c
}
\toprule
\textbf{ID} & \textbf{Title} & \textbf{Publication Date} \\
\midrule

\rowcolor{gray!15}
\multicolumn{3}{l}{\textit{(D6) Health News}} \\
31 & Amid PFAS Fallout, Maine Residents Navigate Medical Risks & 2025-07-29 \\

\midrule
\rowcolor{gray!15}
\multicolumn{3}{l}{\textit{(D7) Wildlife}} \\
32 & In Japan, Conserving the Genetics of a Sacred Deer & 2024-07-05 \\

\midrule
\rowcolor{gray!15}
\multicolumn{3}{l}{\textit{(D8) Architecture \& Hardware}} \\
33 & Evolution of Software Architecture is Costing Us More Energy & 2025-12-11 \\
34 & Creationism and Evolutionism in Embodied Intelligence & 2025-12-09 \\

\midrule
\rowcolor{gray!15}
\multicolumn{3}{l}{\textit{(D9) CS \& Education}} \\
35 & How Can Vibe Coding Transform Programming Education? & 2025-06-23 \\
36 & Your New Role Requires Strategic Thinking\ldots But You're Stuck in the Weeds & 2025-12-16 \\

\midrule
\rowcolor{gray!15}
\multicolumn{3}{l}{\textit{(D10) Prose \& Poem}} \\
37 & A Description of a City Shower & 1710-10-17 \\
38 & The Raven & 1845-01-01 \\
39 & Little Elephant's Christmas & 1938-01-01 \\
40 & The Tale of Kitty-in-Boots & 2003-01-01 \\

\midrule
\rowcolor{gray!15}
\multicolumn{3}{l}{\textit{(D11) Short Fiction}} \\
41 & A Baby Tramp & 1931-01-01 \\
42 & Cinderella, or the Little Glass Slipper & 1931-01-01 \\

\midrule
\rowcolor{gray!15}
\multicolumn{3}{l}{\textit{(D12) Culture}} \\
43 & `It's a Moment of Death and Rebirth': The Ancient Monuments Saluting the Winter Solstice & 2025-12-21 \\
44 & `Mr.\ Baird Was So Excited That Words Didn't Come': The Office Worker Who Became the First Person Ever to Appear on TV & 2025-09-29 \\
45 & `We Wanted to Make It Real': How Goodfellas Reinvented the Gangster Film & 2025-09-08 \\
46 & `She Would Have Been Stripped of Practically Everything': The Untold Story of Princess Margaret's Forbidden First Love & 2025-10-27 \\

\midrule
\rowcolor{gray!15}
\multicolumn{3}{l}{\textit{(D13) Travel}} \\
47 & An `Uber of the Alps': The Swiss Ski Resort Reinventing Winter & 2025-12-16 \\
48 & How a Rockette Spends Christmas in New York City & 2025-12-17 \\
49 & A Fashion Expert's Insider Guide to Shopping in New York City & 2025-12-04 \\
50 & Seven Countries, One Winner: The Best Christmas Market in Europe & 2025-11-17 \\

\midrule
\rowcolor{gray!15}
\multicolumn{3}{l}{\textit{(D14) Psychology \& Philosophy}} \\
51 & Why Writing Is One of the Best Exercises for Thinking & 2023-11-24 \\
52 & Most People Think They Are Thinking---But They're Really Just Reorganizing Their Biases & 2023-09-26 \\
53 & Logical Language Should Be Your First Foreign Language & 2023-08-22 \\
54 & How to Become a High-Level Conversationalist & 2023-11-14 \\
55 & Why Are Truly Interesting People So Rare? & 2023-11-04 \\

\midrule
\rowcolor{gray!15}
\multicolumn{3}{l}{\textit{(D15) Miscellaneous Talks}} \\
56 & What's the Difference Between People Who Read Regularly and Those Who Don't? & 2022-06-24 \\
57 & Career Advice for Students Majoring in the Humanities & 2023-11-04 \\
58 & Memory, Models, and Robots & 2023-11-04 \\

\midrule
\rowcolor{gray!15}
\multicolumn{3}{l}{\textit{(D16) Tweet}} \\
59 & Every Firefly Is a Torch, and Every Path Leads Somewhere & 2025-12-30 \\
60 & Huangshan: A Millennium of Culture and Poetic Landscapes & 2022-12-08 \\

\bottomrule
\end{tabularx}
\end{table*}

\begin{table*}[t]
\centering
\setlength{\tabcolsep}{3pt} 

\caption{Results of GPT-5.1 under L1 and L2 settings across constraint domains.
(D1) Technique Report, (D2) Paper Reading, (D3) Survey Report, (D4) Popular Science Article, (D5) Principle Analysis, (D6) Healthy News, (D7) Wildlife, (D8) Architecture \& Hardware , (D9) CS \& Education , (D10) Prose Poem Collection, (D11) Short Fiction, (D12) Culture , (D13) Travel, (D14) Psychology \& Philosophy, (D15) Miscellaneous Talks, (D16) Tweet.
}

\adjustbox{max width=\textwidth}{

\begin{tabular}{lllcccccccccccccccc}
\toprule
\multirow{2}{*}{\textbf{Model}} &
\multirow{2}{*}{\textbf{Mode}} &
\multirow{2}{*}{\textbf{Overall}} &
\multicolumn{5}{c}{\textbf{Scientific Domain}} &
\multicolumn{11}{c}{\textbf{Creative Domain}} \\
\cmidrule(lr){4-8}\cmidrule(lr){9-19}
& & &
\textbf{D1} & \textbf{D2} & \textbf{D3} & \textbf{D4} & \textbf{D5} & \textbf{D6} &
\textbf{D7} & \textbf{D8} & \textbf{D9} & \textbf{D10} & \textbf{D11} &
\textbf{D12} & \textbf{D13} & \textbf{D14} & \textbf{D15} & \textbf{D16} \\

\midrule
\rowcolor{gray!15}
\multicolumn{19}{l}{\textit{HAR}} \\

\multirow{2}{*}{GPT-5.1} & 
L1 & 17.59  & 16.94 & 29.23 & 12.76 & 20.37 & 26.77 & 
0.16 & 4.76 & 22.73 & 33.90 & 18.64 & 6.56 & 6.78 & 3.16 & 12.37 & 10.53 & 7.50
\\
& L2  & 19.61  &  21.77 & 29.23  &  15.64 & 19.75 & 30.31  & 0.56
 & 14.29 & 22.73  &  27.12 & 19.49  & 14.75 & 10.17 & 3.16 & 9.28  & 10.53 &  27.50
\\

\midrule

\multirow{2}{*}{Gemini-2.5-Pro} & 
L1 &  14.17 & 14.52 & 22.68 & 9.88 & 14.81 & 14.96 & 
5.26 & 9.52 & 18.18 & 16.95 & 29.66 & 4.92 & 7.63 & 4.21 & 13.40 & 5.26 & 10.00
\\
& L2  &  17.89 & 21.77 & 24.10 & 11.11 & 17.90 & 22.44 & 
5.26 & 9.52 & 15.91 & 15.25 & 30.51 & 11.48 & 10.17 & 9.47 & 20.62 & 18.42 & 12.50
\\
\midrule
\multirow{2}{*}{Gemini-2.5-flash} & 
L1 &  13.39 & 14.52 & 21.03 & 6.17 & 14.20 & 18.90 & 
10.53 & 9.52 & 15.91 & 25.42 & 18.64 & 8.20 & 8.47 & 2.11 & 13.40 & 2.63 & 5.00 

\\
& L2  &  16.05 & 17.74 & 26.15 & 11.52 & 16.67 & 26.77 & 
0.56 & 4.76 & 18.18 & 8.47 & 26.27 & 11.48 & 0.85 & 2.11 & 12.37 & 5.26 & 15.00 

\\

\midrule

\multirow{2}{*}{DeepSeek-V3.2} & 
L1 & 6.99 & 2.42 & 14.87 & 4.53 & 6.17 & 10.63 & 
0.56 & 1.16 & 6.82 & 11.86 & 13.56 & 6.56 & 2.54 & 2.11 & 1.03 & 2.63 & 2.50 

\\
& L2  & 10.37 & 11.29 & 16.41 & 9.05 & 10.49 & 16.14 & 
0.56 & 4.76 & 11.36 & 13.56 & 19.49 & 1.64 & 0.42 & 2.11 & 4.12 & 1.23 & 12.50

\\

\midrule
\rowcolor{gray!15}
\multicolumn{19}{l}{\textit{KED}} \\

\multirow{2}{*}{GPT-5.1} & 
L1 & 11.38  & 6.75  & 10.60  & 15.61  & 14.13  &  13.14 & 0.00
  & 10.00   & 14.60  & 12.41  & 1.40  &  13.00 &  10.83 &  12.00   & 7.50  & 10.50  &  11.60
\\
& L2  & 11.16 & 7.28   &  11.36  &  13.37 & 16.04  &  12.86 & 
 12.00 & 9.00   & 14.00  &  12.53  & 2.20   &  11.57 & 11.25  & 7.00  & 8.25  &  8.67 &  7.00 
\\

\midrule

\multirow{2}{*}{Gemini-2.5-Pro} & 
L1 &  9.17 &  6.94 &  8.95 & 11.42  &  10.88 & 15.17  & 10.00
  &  9.50 &  7.57 & 14.25  & 0.50   &  16.00 & 7.89  & 7.50  & 7.31  &  18.00 &  11.50
\\
& L2  & 9.39 &  5.04 &  10.54 & 9.70  & 11.72  & 13.94  & 29.00 
  &  6.00 & 9.43   &  11.62 &  0.81 &  9.14 &  8.17 & 9.25   &  8.10 & 15.43  &  6.80
\\
\midrule
\multirow{2}{*}{Gemini-2.5-flash} & 
L1 &  9.68 & 5.24  & 9.24  & 12.29  & 13.80  & 13.04  & 11.00
  & 8.00  & 12.86  & 10.00  & 1.00   & 9.00  & 12.10  &  7.00 & 6.10  &  8.00 &  5.50

\\
& L2  & 10.24 &  7.60 & 9.47  & 13.96  & 12.00  &  13.74 & 0.00
  &  12.00 & 13.25   & 9.20   & 1.52 &  9.71 & 10.00 &  11.00 &  7.50 & 17.00   &  7.67 

\\

\midrule

\multirow{2}{*}{DeepSeek-V3.2} & 
L1 & 10.18 &  2.33 &  11.43 &  10.30 &  10.70 & 14.00  & 0.00
  & 0.00  &  10.33 & 10.57  &  2.31 & 9.75   & 7.33   & 16.50  & 7.00  & 16.00  &  11.00

\\
& L2  & 9.97 & 8.57  &  9.91 & 14.29  &  11.69 & 12.00 & 0.00
  &  14.00 &  8.86 & 7.00  &  1.90 & 9.86 & 9.00  &  7.50  & 9.17  & 19.00  &  7.83

\\

\bottomrule
\end{tabular}
}

\label{tab:gpt51_constraint_domains}
\end{table*}

\begin{table*}[t]
\centering
\begin{tcolorbox}[
  width=\textwidth,
  colback=white,
  colframe=black!40,
  boxrule=0.6pt,
  arc=3pt,
  left=10pt,
  right=10pt,
  top=8pt,
  bottom=8pt
]

\textbf{Instructions.} \\
You have just completed a writing task with the assistant.
Please rate the following statements based on your experience in this interaction.

Use a \textbf{7-point Likert scale},
where \emph{1 = Very Bad }
and \emph{ 7 = Very Good}.

\vspace{0.6em}
\hrule
\vspace{0.8em}

\subsection*{A. Perceived Workload}

\begin{itemize}[leftmargin=1.2em]
  \item \textbf{Q1. Mental Demand} \\
  How mentally demanding was the writing task under this interaction setting?

  \item \textbf{Q2. Effort} \\
  How much effort did you have to invest to complete the writing task?

  \item \textbf{Q3. Temporal Demand} \\
  How much time pressure did you feel while writing with the assistant?

  \item \textbf{Q4. Frustration} \\
  How frustrated, irritated, or annoyed did you feel during the interaction?
\end{itemize}

\vspace{0.8em}
\hrule
\vspace{0.8em}

\subsection*{B. Usefulness}

\begin{itemize}[leftmargin=1.2em]
  \item \textbf{Q5. Intent Alignment} \\
  The suggestion matched what I was trying to express at that moment.

  \item \textbf{Q6. Decision Facilitation} \\
  The suggestion helped me decide how to continue writing.
\end{itemize}

\vspace{0.8em}
\hrule
\vspace{0.8em}

\subsection*{C. Customization}

\begin{itemize}[leftmargin=1.2em]
  \item \textbf{Q7. Style Understanding} \\
  The assistant understood my writing style and preferences.

  \item \textbf{Q8. Alignment Over Time} \\
  The assistant became more aligned with my intent as the interaction progressed.
\end{itemize}

\vspace{0.8em}
\hrule
\vspace{0.8em}

\subsection*{D. Adaptability}

\begin{itemize}[leftmargin=1.2em]
  \item \textbf{Q9. Timing Appropriateness} \\
  Suggestions were provided at appropriate moments during writing.

  \item \textbf{Q10. Sense of Control} \\
  I felt in control of when and how the assistant intervened.
\end{itemize}

\end{tcolorbox}
\caption{Post-task questionnaire for the user study, covering Perceived Workload, Usefulness, Customization, and Adaptability.
All items are rated on a 7-point Likert scale from \emph{Very Bad} (1) to \emph{Very Good} (7).}
\label{tab:user_study_questionnaire}
\end{table*}

\begin{table*}[t]
\centering
\setlength{\tabcolsep}{3pt}

\caption{User Study Full Score}

\adjustbox{max width=\textwidth}{
\begin{tabular}{lllcccccccccccccccc}
\toprule
\multirow{2}{*}{\textbf{Interaction Paradigm}} &
\multirow{2}{*}{\textbf{Scale Value}} &
\multicolumn{4}{c}{\textbf{Perceived Workload}} &
\multicolumn{2}{c}{\textbf{Usefulness}} &
\multicolumn{2}{c}{\textbf{Customization}} &
\multicolumn{2}{c}{\textbf{Adaptability}} \\
\cmidrule(lr){3-6}\cmidrule(lr){7-8} \cmidrule(lr){9-10} \cmidrule(lr){11-12}
& &
\textbf{Q1} & \textbf{Q2} & \textbf{Q3} & \textbf{Q4} &
\textbf{Q5} & \textbf{Q6} &
\textbf{Q7} & \textbf{Q8} &
\textbf{Q9} & \textbf{Q10} \\
\midrule

\multirow{7}{*}{\makecell[l]{(L0) User-Initiated \\ Collaboration} } &
1 (Very Bad) & 6 & 5 & 5 & 3 & 3 & 6 & 4 & 2 & 3 & 1 \\
& 2 (Bad) & 5 & 6 & 9 & 10 & 8 & 4 & 7 & 9 & 7 & 12 \\
& 3 (Lightly Bad) & 4 & 9 & 6 & 6 & 5 & 4 & 3 & 3 & 5 & 4 \\
& 4 (Moderate) & 13 & 10 & 7 & 9 & 11 & 14 & 11 & 13 & 11 & 10 \\
& 5 (Lightly Good) & 2 & 0 & 3 & 1 & 2 & 0 & 2 & 1 & 2 & 1 \\
& 6 (Good) & 0 & 0 & 0 & 1 & 1 & 2 & 3 & 2 & 2 & 2 \\
& 7 (Very Good) & 0 & 0 & 0 & 0 & 0 & 0 & 0 & 0 & 0 & 0 \\
\midrule

\multirow{7}{*}{
\makecell[l]{(L1) Stateless Proactive \\ Collaboration}} &
1 (Very Bad) & 0 & 0 & 3 & 2 & 0 & 2 & 1 & 0 & 3 & 4 \\
& 2 (Bad) & 0 & 3 & 2 & 0 & 1 & 2 & 4 & 4 & 3 & 2 \\
& 3 (Lightly Bad) & 8 & 6 & 4 & 9 & 3 & 3 & 12 & 7 & 10 & 7 \\
& 4 (Moderate) & 10 & 9 & 11 & 9 & 11 & 9 & 4 & 6 & 5 & 6 \\
& 5 (Lightly Good) & 4 & 6 & 5 & 5 & 6 & 7 & 6 & 9 & 6 & 7 \\
& 6 (Good) & 7 & 5 & 4 & 4 & 7 & 4 & 2 & 2 & 3 & 3 \\
& 7 (Very Good) & 1 & 1 & 1 & 1 & 2 & 3 & 1 & 2 & 0 & 1 \\
\midrule

\multirow{7}{*}{
\makecell[l]{(L2)Stateful Proactive  \\ Collaboration}} &
1 (Very Bad) & 0 & 1 & 0 & 0 & 0 & 0 & 0 & 0 & 1 & 0 \\
& 2 (Bad) & 1 & 1 & 4 & 1 & 0 & 3 & 1 & 1 & 3 & 1 \\
& 3 (Lightly Bad) & 2 & 1 & 2 & 3 & 4 & 4 & 3 & 3 & 2 & 5 \\
& 4 (Moderate) & 7 & 8 & 5 & 5 & 10 & 6 & 6 & 6 & 3 & 8 \\
& 5 (Lightly Good) & 10 & 5 & 8 & 12 & 5 & 6 & 12 & 8 & 9 & 9 \\
& 6 (Good) & 7 & 7 & 6 & 7 & 7 & 6 & 7 & 6 & 6 & 4 \\
& 7 (Very Good) & 3 & 7 & 5 & 2 & 4 & 5 & 1 & 6 & 6 & 3 \\
\midrule

\multirow{7}{*}{
\makecell[l]{(L3) Adaptive Proactive \\ Collaboration}} &
1 (Very Bad) & 2 & 0 & 0 & 0 & 1 & 0 & 1 & 1 & 1 & 0 \\
& 2 (Bad) & 0 & 1 & 1 & 0 & 3 & 1 & 1 & 0 & 1 & 1 \\
& 3 (Lightly Bad) & 2 & 4 & 4 & 6 & 1 & 4 & 2 & 6 & 2 & 2 \\
& 4 (Moderate) & 6 & 3 & 3 & 5 & 2 & 8 & 5 & 2 & 5 & 6 \\
& 5 (Lightly Good) & 10 & 9 & 9 & 6 & 10 & 4 & 9 & 9 & 12 & 9 \\
& 6 (Good) & 3 & 7 & 6 & 7 & 6 & 6 & 5 & 6 & 5 & 5 \\
& 7 (Very Good) & 7 & 6 & 7 & 6 & 7 & 7 & 7 & 6 & 4 & 7 \\
\bottomrule
\end{tabular}
}

\label{tab:user_study_scores_full}
\end{table*}

\onecolumn
\begin{tcolorbox}[HAR]

\textbf{Task Description}

You are a professional researcher collaborating with your Writing Assistant to complete a document. The Writing Assistant will fill in subsequent content based on the user's typed input \texttt{<USER\_INPUT>}. Your responsibility is: given the user's typed input \texttt{<USER\_INPUT>}, the Assistant-generated completion \texttt{<COMPLETION>}, and the reference text (original content at the completion position) \texttt{<REFERENCE>}, determine whether this completion should be \texttt{"accepted"} or \texttt{"rejected"}.

\textbf{Decision Criteria (Checklist)}

Please execute strictly in order, first checking blocking conditions. When triggered, make a direct judgment without proceeding to subsequent comparisons.

\begin{enumerate}
    \item \textbf{Start repetition:} if \texttt{<COMPLETION>} starts with a \textbf{repetition} of \texttt{<USER\_INPUT>} $\rightarrow$ reject.
    \item \textbf{Language mismatch:} if \texttt{<COMPLETION>} and \texttt{<REFERENCE>} are in different languages (e.g., English vs.\ Simplified Chinese vs.\ Traditional Chinese) $\rightarrow$ reject.
    \item \textbf{Semantic coherence:} if \texttt{<USER\_INPUT>} + \texttt{<COMPLETION>} form a contradictory or incoherent sequence $\rightarrow$ reject.
    \item \textbf{Early semantic overlap with \texttt{<REFERENCE>}:} if the overlapping content between the beginning of \texttt{<COMPLETION>} and the beginning of \texttt{<REFERENCE>} accounts for more than 50\% of the total \texttt{<COMPLETION>} content $\rightarrow$ accept.
    \item \textbf{Paired punctuation mark closure:} unclosed quotes/brackets introduced in \texttt{<USER\_INPUT>} remain unclosed in \texttt{<COMPLETION>} $\rightarrow$ reject.
    \item \textbf{Markdown/LaTeX/Code closure:} any opened Markdown/LaTeX/code fences from \texttt{<USER\_INPUT>} not properly closed in \texttt{<COMPLETION>} $\rightarrow$ reject.
    \item \textbf{Format mismatch:} if format mismatch between \texttt{<COMPLETION>} and \texttt{<REFERENCE>} (headings, tables, lists, text) $\rightarrow$ reject.
    \item \textbf{Format consistency with preceding text:} the format/style used in \texttt{<COMPLETION>} diverges from the format established by the preceding content of \texttt{<USER\_INPUT>} and \texttt{<REFERENCE>}, e.g., changes in list type, indentation levels, heading levels $\rightarrow$ reject.
    \item \textbf{Depth mismatch:} specificity/level of detail diverges from \texttt{<REFERENCE>} beyond tolerance ($\pm 30\%$) $\rightarrow$ reject.
    \item \textbf{Style/Register mismatch:} academic vs.\ conversational vs.\ authoritative diverges from \texttt{<REFERENCE>} $\rightarrow$ reject.
    \item \textbf{Sentence type mismatch:} declarative vs.\ interrogative vs.\ imperative diverges from \texttt{<REFERENCE>} $\rightarrow$ reject.
    \item \textbf{Personal perspective check:} If \texttt{<COMPLETION>} shifts the narrative perspective (first person, second person, third person) established in \texttt{<USER\_INPUT>} $\rightarrow$ reject.
    \item \textbf{Subset acceptance (lists/tables):} if \texttt{<COMPLETION>} is a true subset of \texttt{<REFERENCE>} (same order, no new items; tables must keep columns and header order) $\rightarrow$ accept.
    \item \textbf{Topic divergence:} if \texttt{<COMPLETION>} and \texttt{<REFERENCE>} address different topics $\rightarrow$ reject.
    \item \textbf{Key entities:} missing or altered key entities (names, trial IDs, datasets, metrics, units, dates, statistics data, terminology) compared to \texttt{<REFERENCE>} $\rightarrow$ reject.
    \item \textbf{Intent mismatch:} if the intent of \texttt{<COMPLETION>} differs from \texttt{<REFERENCE>} (e.g., summary vs.\ argument vs.\ instruction) $\rightarrow$ reject.
    \item \textbf{Comprehensive judgment:} if none of the above conditions are triggered, please carefully analyze the \texttt{<USER\_INPUT>} and \texttt{<REFERENCE>}, then make a comprehensive judgment on whether to accept the \texttt{<COMPLETION>} based on style, semantics, entities, and other factors.
\end{enumerate}

\textbf{Output Format}

Please strictly use the following JSON format for output and enclose the entire JSON in \verb|\boxed{ ... }|,

\begin{lstlisting}[language=json,basicstyle=\ttfamily\small,breaklines=true,breakatwhitespace=true]
\boxed{
  "accept": true | false,
  "triggered_condition": "rule number and name (e.g., \"1. Start repetition\")",
  "reasoning": "Provide your reasoning, explicitly mentioning which rule number was triggered and why."
}
\end{lstlisting}

\textbf{Task Input}

\begin{verbatim}
<USER_INPUT>:"{context}"
<REFERENCE>: "{sentence_A}"
<COMPLETION>: "{sentence_B}"
\end{verbatim}

\end{tcolorbox}

\begin{tcolorbox}[logic]

\textbf{System Prompt}

You are a discourse analysis expert. Your task is to identify the \textbf{Core Backbone Logical Relationship} of a complex sentence and determine whether the backbone logic of two sentences is consistent.

The backbone logic refers to the primary logical relationship that drives the main proposition of the sentence. You should focus on the relationship between main clauses and ignore modifying, explanatory, or subsidiary logic (e.g., parenthetical remarks, citations, or examples).

\textbf{Template Description}

The input consists of two sentences, labeled as \texttt{Preference} and \texttt{Prediction}:

\begin{verbatim}
Preference: {sentence_A}

Prediction: {sentence_B}
\end{verbatim}

\textbf{Guidelines}

\begin{enumerate}
    \item Identify the \textbf{Core Backbone Logical Relationship} for each sentence independently.
    
    \item The backbone logic should reflect the relationship connecting the main propositions (Main Clause to Main Clause), rather than local or modifying relationships.
    
    \item Secondary or subsidiary relationships (e.g., elaboration, supplements, parenthetical comments) should \textbf{not} be treated as the backbone logic.
    
    \item When the logical relationship is ambiguous, follow the priority hierarchy below:
    \begin{itemize}
        \item Relationship between Main Clauses $>$ Relationship between Main and Subordinate Clauses $>$ Modifying Logic.
        \item Logic that drives semantic progression (Causal, Contrast, Sequence, Temporal) $>$ Expansion or Elaboration logic.
        \item Explicit discourse markers (e.g., ``because'', ``so'', ``but'', ``and'', ``subsequently'', ``firstly'') usually indicate the backbone logic.
    \end{itemize}
\end{enumerate}

\textbf{Optional Logic Labels}

\begin{enumerate}
    \item Sequence / Connection
    \item Causal (including Purpose)
    \item Contrast / Concession
    \item Elaboration / Explanation
    \item Temporal
    \item Frame / Organization
    \item Other
\end{enumerate}

\textbf{Tasks}

\begin{enumerate}
    \item Assign a backbone logic label ID to the \texttt{Preference} sentence.
    \item Assign a backbone logic label ID to the \texttt{Prediction} sentence.
    \item Determine whether the backbone logical relationships of the two sentences are the same or different.
\end{enumerate}

\textbf{Output Format}

Please return the result in \textbf{JSON format}, wrapped entirely in \verb|\boxed{}|. Do not include any additional text.

\begin{lstlisting}[breaklines=true, basicstyle=\ttfamily, columns=fullflexible]
```json
\boxed{
  "Preference_logic": <ID_Number>,
  "Prediction_logic": <ID_Number>,
  "logicalCompare": "A/B"  // Use "A" for Same, "B" for Different
}
\end{lstlisting}

\end{tcolorbox}

\begin{tcolorbox}[stylistic]

  \textbf{System Prompt}

  You are a professional linguistic style analyst tasked with evaluating the style similarity between two sentences. Please conduct a detailed analysis based on the following criteria.

  \textbf{Evaluation Requirements}

  Please analyze the style similarity of the two sentences from the following four dimensions and provide a score of 1–10. After scoring each dimension, calculate the overall similarity according to the weighted percentages:

  \begin{enumerate}
    \item \textbf{Lexical Style Analysis (Weight 25\%)}
    \begin{itemize}
        \item Formality Contrast (Written vs. Colloquial).
        \item Part-of-Speech Features (Pronouns, emotive words, abbreviations, conjunctions).
        \item Score Range: 1–10
    \end{itemize}
    
    \item \textbf{Syntactic Structure Analysis (Weight 25\%)}
    \begin{itemize}
        \item Sentence Length (Comparison of average word count).
        \item Syntactic Complexity (Clauses, passive voice, compound/complex structures).
        \item Score Range: 1–10
    \end{itemize}

    \item \textbf{Linguistic Style Features (Weight 30\%)}
    \begin{itemize}
        \item Lexical Richness (Diversity and repetition patterns).
        \item Rhetorical Devices (Metaphors, parallelism, etc.).
        \item Tone and Mood (Emotional coloring, sentence types).
        \item Score Range: 1–10
    \end{itemize}

    \item \textbf{Genre/Stylistic Features (Weight 20\%)}
    \begin{itemize}
        \item Register Consistency (Matching of formality and professional level).
        \item Expression Habits (Logical organization and methods of emphasis).
        \item Score Range: 1–10
    \end{itemize}
  \end{enumerate}

  \textbf{Output Format (Strictly follow JSON structure)}

  Please wrap the complete JSON response in \texttt{\textbackslash boxed\{\{\}\}} and do not include any other text.

\begin{verbatim}
\boxed{{
  "analysis": {{
    "lexical_style": {{
      "score1": 0,
      "comment": "<Analysis>"
    }},
    "syntactic_structure": {{
      "score2": 0,
      "comment": "<Analysis>"
    }},
    "linguistic_features": {{
      "score3": 0,
      "comment": "<Analysis>"
    }},
    "genre_features": {{
      "score4": 0,
      "comment": "<Analysis>"
    }}
  }},
  "styleSimilarity_overall": 0.0,
  "conclusion": "<Analysis>"
}}
\end{verbatim}

  \textbf{Important Requirements}

  \begin{itemize}
    \item Wrap the complete JSON output in \texttt{\textbackslash boxed\{\{\}\}}.
    \item Do not include any explanatory text or extra content.
    \item Ensure the JSON format is completely correct.
  \end{itemize}

  \textbf{Weighted Calculation Rule}

  Round the result to one decimal place, within the range of 1–10.

\begin{verbatim}
styleSimilarity_overall = (lexical_style.score1 * 0.25 + 
                           syntactic_structure.score2 * 0.25 + 
                           linguistic_features.score3 * 0.30 + 
                           genre_features.score4 * 0.20)
\end{verbatim}

  \textbf{Sentences to Analyze}

  Sentence A: \{sentence\_A\}
  
  Sentence B: \{sentence\_B\}
  
\end{tcolorbox}

\begin{tcolorbox}[semantic]

  \textbf{Task}

  You are a semantic similarity evaluator. Given two sentences, please score them between 0 and 10 (integers only): 0 indicates completely irrelevant or contradictory, and 10 indicates semantic equivalence (mere paraphrasing with almost no difference). Focus on the "meaning" rather than surface wording.

  \textbf{Instructions}

  Focus on the core propositional content (Who/What/When/Where/Why/How). Ignore style, tone, punctuation, and minor word order differences.

  Deduct points if facts, scope, negation/affirmation, quantity, time, location, or modality (can/must/may) differ.

  \textbf{Distinctions}

  \begin{itemize}
    \item Entailment (A entails B or B entails A) $\to$ Usually 7--9, depending on the missing details.
    \item Paraphrasing / Bidirectional Entailment $\to$ 9--10.
    \item Same topic but different claims/assertions $\to$ 3--5.
    \item Contradictory or mutually exclusive facts $\to$ 0--2.
    \item If one sentence is more specific, moderately deduct points based on whether these extra details are critical to the claim.
    \item If semantics are uncertain, conservatively choose the lower score.
    \item Do not use external knowledge beyond common sense.
    \item Wrap the complete JSON output in \texttt{\textbackslash boxed\{\{\}\}}.
  \end{itemize}

  \textbf{Scoring Reference (Non-absolute)}

  \begin{itemize}
    \item 0--1: Irrelevant or directly contradictory.
    \item 2: Basically irrelevant, only weak lexical overlap.
    \item 3--4: Same broad topic, but different assertions.
    \item 5--6: Overlap exists, but with significant differences (time/quantity/polarity/scope).
    \item 7--8: High overlap or one-way entailment, with only minor omissions.
    \item 9: Almost synonymous paraphrasing, with only trivial differences.
    \item 10: Completely equivalent paraphrasing.
  \end{itemize}

  \textbf{Output Format}

\begin{verbatim}
```json
\boxed{{
 "semanticSimilarity_score": <Integer between 0-10>,
 "reason": "<1–2 sentences brief reason>",
}}

\end{verbatim}

\textbf{Evaluate Now}

\begin{verbatim}
Sentence 1: {sentence_A}

Sentence 2: {sentence_B}

Return JSON only.
\end{verbatim}

\end{tcolorbox}

\begin{tcolorbox}[holistic]

\textbf{System Prompt}

You are an academic evaluator specializing in semantic and pragmatic comparison.  
Your task is to compare two sentences and determine whether the \texttt{Prediction} is sufficiently similar to the \texttt{Preference} to be accepted as expressing the same meaning.

\textbf{Task Description}

You need to assess the similarity between the given \texttt{Preference} sentence and the \texttt{Prediction} sentence from four complementary dimensions.  
Each dimension should be scored on a scale from 1 to 10, where 1 indicates complete dissimilarity and 10 indicates near-identical equivalence.

\textbf{Similarity Dimensions}

\begin{enumerate}
    \item \textbf{Entity Similarity}:  
    Evaluate whether the entities involved in both sentences (e.g., names, locations, organizations, or other proper nouns) are consistent, aligned, or closely related.
    
    \item \textbf{Logical Similarity}:  
    Assess whether the logical structure, causal relations, and reasoning flow of the two sentences are consistent.
    
    \item \textbf{Style Similarity}:  
    Determine whether the language style, tone, rhetorical manner, and formality level are similar.
    
    \item \textbf{Semantic Similarity}:  
    Judge whether the overall meaning, core intent, and semantic content of the two sentences are equivalent.
\end{enumerate}

\textbf{Final Decision Rule}

After scoring and providing reasoning for all four dimensions, synthesize them into an overall judgment:

\begin{itemize}
    \item If the \texttt{Prediction} and \texttt{Preference} are overall the same or highly consistent, set \texttt{"accept"} to \texttt{true}.
    \item If there are clear or substantial differences, set \texttt{"accept"} to \texttt{false}.
\end{itemize}

\textbf{Output Format}

You must strictly follow the JSON schema below and enclose the entire output within \texttt{\textbackslash boxed\{\dots\}}.

\begin{verbatim}
\boxed{
{
  "similarities": [
    {
      "aspect": "Entity Similarity",
      "score1": <Integer between 1 and 10>,
      "reasoning1": "<Reasoning for score>"
    },
    {
      "aspect": "Logical Similarity",
      "score2": <Integer between 1 and 10>,
      "reasoning2": "<Reasoning for score>"
    },
    {
      "aspect": "Style Similarity",
      "score3": <Integer between 1 and 10>,
      "reasoning3": "<Reasoning for score>"
    },
    {
      "aspect": "Semantic Similarity",
      "score4": <Integer between 1 and 10>,
      "reasoning4": "<Reasoning for score>"
    }
  ],
  "accept": true | false,
  "overall_reasoning": "<Final reasoning after synthesizing the four dimensions>"
}
}
\end{verbatim}

\textbf{Sentences to Compare}

\begin{verbatim}
Preference: "{sentence_A}"
Prediction: "{sentence_B}"
\end{verbatim}

\end{tcolorbox}

\begin{tcolorbox}[edit_distance]

\textbf{System Prompt}

You are a professional researcher collaborating with your Writing Assistant to complete a document. The Writing Assistant will generate subsequent text \texttt{<COMPLETION>} based on the user's input \texttt{<USER\_INPUT>}. Your responsibility is to evaluate the editing cost required to transform the Assistant-generated \texttt{<COMPLETION>} into the \texttt{<REFERENCE>} text.  
To achieve this, you must first holistically assess the differences between the two texts and then generate a clear \textbf{List of Editing Actions} given predefined actions. The final \textbf{Total Editing Cost} is the sum of the scores for all actions.

\textbf{Actions}

\begin{enumerate}
    \item \texttt{"ADD"}
    \item \texttt{"DELETE"}
    \item \texttt{"MODIFY"}
\end{enumerate}

\textbf{Cost Assessment Rules}

The cost of each action correlates with the \textbf{entities} involved and the \textbf{transitional phrasing}.  
Generally, complex entities incur higher costs than simple entities, and substantial transitional phrasing revisions are more costly than minor ones.

\textbf{Cost Assessment of Entities}
\begin{itemize}
    \item A \textbf{complex entity} in action \texttt{"ADD"} or \texttt{"MODIFY"} is scored as \textbf{3 points}.
    \item A \textbf{simple entity} in action \texttt{"ADD"} or \texttt{"MODIFY"} is scored as \textbf{1 point}.
    \item Action \texttt{"DELETE"} is assessed as part of the transitional phrasing evaluation.
\end{itemize}

\textbf{Cost Assessment of Transitional Phrasing}
\begin{itemize}
    \item Score \textbf{0} point if the relational description is identical or semantically equivalent.
    \item Score \textbf{1} point for minor, simple additions, deletions, or modifications of words.
    \item Score \textbf{2} points for substantial modifications to the core verbs, phrases, or overall structure of the relational description.
\end{itemize}

\textbf{Complexity Assessment of Entities}

\textbf{Complex Entity (COMPLEX)}:  
Entities that require fact-checking, external knowledge, or precise recall. These typically include proper nouns (e.g., names of people, places, organizations), scientific, technical, or legal terminology, specific product or substance names, and precise numbers, models, or dates.

\textbf{Simple Entity (SIMPLE)}:  
Common nouns, descriptive words, or common-sense concepts that can be intuitively and quickly verified or modified.

\textbf{Assessment Procedure}

Generate a list of editing actions. The list of editing instructions for the \texttt{<COMPLETION>} may consist of various combinations of actions.  
Intuitively, you prefer action sequences with lower cost. The cost is computed as follows:

\begin{enumerate}
    \item \textbf{Single Action Cost Accumulation}:  
    The cost for each action is the cumulative sum of the scores of all involved entities and the transitional phrasing complexity score.
    \item \textbf{Total Action Cost Accumulation}:  
    The final \textbf{[Total Editing Cost]} is calculated by summing the costs of all actions.
\end{enumerate}

\textbf{Input and Output Example}

\textbf{Input}
\begin{lstlisting}[breaklines=true, basicstyle=\ttfamily, columns=fullflexible]
<REFERENCE>: "The protein is eluted from the polyacrylamide gel and immobilized on the membrane surface."
<COMPLETION>: "The protein is transferred from the gel to the membrane."
\end{lstlisting}

\textbf{Output}

Please strictly use the following JSON format for the output and enclose the entire JSON object within a \texttt{\textbackslash boxed\{\dots\}} block:

\begin{lstlisting}[breaklines=true, basicstyle=\ttfamily, columns=fullflexible]
\boxed{
  {
    "edit_plan": [
      {
        "operation": "MODIFY",
        "instruction": "Revise the entity 'protein' to 'The protein'.",
        "cost": 1,
        "reasoning": "Modification of a single simple entity."
      },
      {
        "operation": "MODIFY",
        "instruction": "Revise the entity 'gel' to 'polyacrylamide gel'.",
        "cost": 3,
        "reasoning": "Involves the modification of one complex entity."
      },
      {
        "operation": "MODIFY",
        "instruction": "Revise the phrasing 'is transferred... to' to 'is eluted from... and immobilized on the... surface'.",
        "cost": 2,
        "reasoning": "Substantial modification to the core verb and overall structure."
      }
    ],
    "total_editing_cost": 6,
    "summary": "The total editing cost is 6 points, comprising the modification of one complex entity and one simple entity, along with a substantial revision to the transitional phrasing."
  }
}
\end{lstlisting}

\textbf{Task Input}

\begin{verbatim}
<REFERENCE>: "{sentence_A}"
<COMPLETION>: "{sentence_B}"
\end{verbatim}

\end{tcolorbox}

\begin{tcolorbox}[coherence_train]

\textbf{System Prompt}

You are a professional researcher collaborating with your Writing Assistant to complete a document.  
The Writing Assistant will fill in subsequent content based on the user's typed input \texttt{<USER\_INPUT>}.  
Your responsibility is: given the user's typed input \texttt{<USER\_INPUT>} and the Assistant-generated completion \texttt{<COMPLETION>}, determine whether this completion is a coherent, contextually appropriate, and logically consistent extension of a given prefix.

\textbf{Decision Criteria (Checklist)}

\begin{enumerate}
    \item \texttt{<COMPLETION>} must logically follow from the \texttt{<USER\_INPUT>}.
    \item \texttt{<COMPLETION>} must maintain the same topic, style, and tone.
    \item \texttt{<COMPLETION>} must smoothly continue the narrative or argument flow.
    \item \texttt{<COMPLETION>} should feel like a natural next part of the \texttt{<USER\_INPUT>}.
    \item \texttt{<COMPLETION>} must \textbf{directly continue from the \texttt{<USER\_INPUT>}} — the first character of the completion must immediately follow the last character of the \texttt{<USER\_INPUT>}.
    \item When \texttt{<USER\_INPUT>} and \texttt{<COMPLETION>} are seamlessly concatenated, they should form a naturally expressed and logically coherent passage.
\end{enumerate}

\textbf{Evaluation Input}

\begin{verbatim}
<USER_INPUT>: "{context}"
<COMPLETION>: "{predicted}"
\end{verbatim}

\textbf{Scoring Rules}

If the \texttt{<COMPLETION>} satisfies \emph{all} of the above checklist items, assign a score of \textbf{1}.  
If the \texttt{<COMPLETION>} fails to meet \emph{any} of the checklist items, assign a score of \textbf{0}.

\textbf{Output Format}

\begin{verbatim}
Score: 0
or
Score: 1
\end{verbatim}

\end{tcolorbox}

\begin{tcolorbox}[semantic_train]

\textbf{System Prompt}

You are a careful semantic evaluation expert. Your task is to assess whether a predicted sentence is semantically equivalent to a reference paragraph.

\textbf{Task}

Given a \textbf{Predicted} sentence and a \textbf{Reference} paragraph, determine whether the Predicted text achieves at least \textbf{80\% semantic similarity} with the Reference.

The inputs are provided in the following format:

\begin{verbatim}
Reference:
{reference}

Predicted:
{predicted}
\end{verbatim}

\textbf{Scoring Rules}

\begin{itemize}
    \item If the Predicted text expresses the same core meaning or main idea as the Reference, with more than 80\% semantic similarity, assign a score of \texttt{1}.
    \item If the Predicted text shares less than or equal to 80\% semantic similarity with the Reference, or conveys a clearly different meaning, assign a score of \texttt{0}.
    \item The judgment should focus on \emph{semantic overlap} rather than exact lexical or surface-form matching.
\end{itemize}

\textbf{Output Constraints}

\begin{itemize}
    \item Output only the score on a single line.
    \item Do \emph{not} include any explanations, reasoning steps, or additional text.
\end{itemize}

\textbf{Output Format}

\begin{verbatim}
Score: 0
\end{verbatim}

or

\begin{verbatim}
Score: 1
\end{verbatim}

\end{tcolorbox}

\begin{tcolorbox}[completion_l1]

  \textbf{System Prompt}

You are an intelligent writing continuation assistant. Your task is to continue the user's incomplete text by analyzing their intent at the pause point and generating a suggestion that seamlessly aligns with the preceding text in logic, style, and function.

\textbf{VERY IMPORTANT:} If the pause occurs within a sentence, continue seamlessly from the last part without repeating its ending.

\textbf{Instructions}

\begin{enumerate}
    \item \textbf{Function Alignment}: Determine the functional role needed after the pause (e.g., continuing plot, deepening emotion, expanding arguments, providing transitions, introducing dialogue) and ensure the continuation serves that purpose.
    \item \textbf{Seamless Connection}: If pausing mid-sentence, connect directly with the last word or phrase. If pausing at a sentence boundary, ensure structural alignment with the preceding text.
    \item \textbf{Style Mimicry}: Match the author's word choice and sentence structure to maintain their distinctive voice.
    \item \textbf{Goal-Oriented}: Follow the article's narrative or argumentative trajectory, achieving a sense of ``spiritual resemblance'' that organically advances the content.
    \item \textbf{Brevity}: Keep continuations very short—typically one sentence fragment or phrase. Aim for a spark of inspiration rather than elaboration.
\end{enumerate}

\textbf{Input Format}

\begin{verbatim}
CONTEXT:
\end{verbatim}

\end{tcolorbox}

\begin{tcolorbox}[completion_l2]

\textbf{System Prompt}

You are an intelligent continuation assistant named ``Lingxi'' (Consonance), an intuitive creative partner. Your task is to complete the user's unfinished text. By analyzing the user's intent at the point of interruption, generate a continuation suggestion that is logically, stylistically, and functionally a perfect match, seamlessly connected, and integral to the preceding context.

\medskip
\textbf{VERY IMPORTANT:} If the pause occurs within a sentence, the continuation should begin seamlessly from the last part of the preceding text, without repeating its ending.

\medskip
\textbf{Instructions}

\begin{itemize}
    \item \textbf{Functional Alignment:} Deeply analyze the context to determine the functional role required for the \textbf{content to be continued} (e.g., advancing the plot, deepening emotions, expanding an argument, providing a transition, introducing dialogue, etc.), ensuring the continuation aligns functionally with the preceding text.
    
    \item \textbf{Seamless Transition:} If the interruption occurs mid-sentence, the first word or character of the continuation must follow the preceding text seamlessly. If the interruption occurs at the beginning of a sentence, the continuation must be structurally compatible with the preceding text.
    
    \item \textbf{Stylistic Mimicry:} Your diction and phrasing must align with the author's stylistic preferences, ensuring the generated content remains perfectly consistent with the author's ``voice.''
    
    \item \textbf{Goal-Oriented:} Ensure the continuation strategically adheres to the document profile, capturing the ``spirit'' rather than merely the ``form,'' and organically driving the narrative or argumentation forward.
    
    \item \textbf{Conciseness:} The continuation should be extremely brief, aiming to spark inspiration. It should typically be a text fragment no longer than a single sentence; in some cases, a single phrase suffices. Avoid lengthy discourse.
\end{itemize}

\medskip
\textbf{Input Format}

The input consists of a \texttt{document\_snapshot}, which captures the document's current state and the edits made in this round. It is formatted as follows:

\begin{verbatim}
document_snapshot:
    ...unchanged preceding text...
    <del>User-deleted content.</del>
    <accept>Content from your previous generation that the user accepted.</accept>
    <reject>Content from your previous generation that the user rejected.</reject>
    <add>User-added content.</add>
\end{verbatim}

\end{tcolorbox}


\end{document}